\documentclass[twocolumn,showpacs,aps,prl]{revtex4}
\pdfoutput=1
\usepackage{graphicx}
\usepackage{dcolumn}
\usepackage{amsmath}
\usepackage{latexsym}
\usepackage{morefloats}

\begin {document}

\title {Band structure in collective motion with quenched range of interaction}
\author
{
Biplab Bhattacherjee and S. S. Manna
}
\affiliation
{
\begin {tabular}{c}
Satyendra Nath Bose National Centre for Basic Sciences,
Block-JD, Sector-III, Salt Lake, Kolkata-700106, India \\
\end{tabular}
}
\begin{abstract}
      A variant of the well known Vicsek model of the collective motion of a group of agents
   has been studied where the range of interactions are spatially quenched and non-overlapping.
   To define such interactions, the underlying two dimensional space is discretized and is 
   divided into the primitive cells of an imaginary square lattice. At any arbitrary time 
   instant, all agents within one cell mutually interact with one another. Therefore, when an 
   agent crosses the boundary of a cell, and moves to a neighboring cell, only then its influence
   is spread to the adjacent cell. Tuning the strength of the scalar noise $\eta$ it has been observed 
   that the system makes a discontinuous transition from a random diffusive phase to an ordered phase through a
   critical noise strength $\eta_c$ where directed bands with high agent densities appear.
   Unlike the original Vicsek model here a host of different types of bands has been observed
   with different angles of orientation and different wrapping numbers. More interestingly,
   two mutually crossed independent sets of simultaneously moving bands are also observed. A 
   prescription for the detailed characterization of different types of bands have been formulated.
\end{abstract}
\maketitle

\section {1. Introduction}

      Among the systems exhibiting non-equilibrium phase transitions under driven noise, the phenomenon
   of Collective Behavior is a well known example \cite{Vicsekreview,Toner1,Toner2,Blair,Czirok,Szabo}. 
   There are a number of living systems in nature which are the prototypical examples of 
   collective motion, such as bacterial colonies \cite{Benjacob}, insect swarms \cite{Rauch}, bird flocks 
   \cite{Feare}, fish schools \cite{Hubbard} etc. All these systems have the following common 
   characteristics: Each one is a collection of living organisms which are self-propelling and their 
   movements are controlled by the influence of other organisms in their close neighborhoods. It has been
   observed that even such a qualitative description provides a good starting point for a theoretical 
   understanding of the collective motion. In the associated models, the living organisms are referred 
   as `agents' in general.
      
      Over the last several years considerable research work has been done to study the collective motion.
   A good portion of this activity has followed the seminal work by Vicsek et. al. \cite{Vicsek}.
   A range of theoretical, numerical and experimental studies have been done \cite{Toner1,Aldana,Chate,
   Baskaran,Blair,Mishra2}. 
   
      In brief, the Vicsek model may be stated as follows. This model describes the dynamical evolution
   of a collection of agents in the continuum Euclidean space with periodic boundary condition. Each
   individual agent is described as a massless point particle with a given position and velocity direction.
   The speeds of all agents are assumed to be the same and a constant value of it is maintained throughout
   their motion. At every time instant, the direction of motion of each agent is determined by its
   interaction with other agents in the local neighborhood, called the {\it interaction zone}. More specifically,
   an interaction zone is the area within a circle of radius $R$ drawn around each individual agent which
   is refreshed at each instant of time. An agent interacts with all neighbors within this zone including
   itself. As the agent moves the interaction zone also moves with it. Agents are also subjected to noise,
   which alters their chosen directions. In this frame work, the Vicsek model exhibits a dynamical phase
   transition from a disordered and incoherent phase to an ordered and coherent phase as the noise level
   is decreased, or the density of agents increased. At the early stage the nature of transition had been 
   claimed to be continuous.

      A number of variants of the original Vicsek model have been introduced to carry out the analytical  
   and numerical studies of collective motion. For example, Chat\'{e}. et al. \cite{Chate1} questioned the 
   continuous nature of the transition. Their extensive numerical studies indicated that the discontinuous  
   nature of the transition appears for system sizes beyond a certain ``crossover'' size that is independent 
   of the magnitude of the self-propulsion speed of the agents. They introduced the vectorial noise, in 
   contrast to the scaler noise of the original Vicsek model, following the argument that the agents can 
   also make errors while estimating the velocity directions of their neighbors. 
   
      It has also been exhibited that similar phase transition can be observed if the agents interact with
   their certain topological neighbors \cite{BBM}, instead of the neighbors within a fixed range $R$ of the 
   interaction zone. Importance of topological neighbors have been revealed by the experiments performed by 
   the European groups of scientists observing the motion of flocks of starling birds indicated the 
   influence of other starlings at their topological neighborhoods \cite {StarFlag}. 
      
      Apart from these variants of the Vicsek model, a number of other models like binary collision model 
   \cite{Bertin2006}, flocking model with a repulsive term \cite{Chate1, chatepre} and other variants with 
   nematic alignment \cite{Chate-nematics, Chate-nematics1, Chate-variants} have also been studied.

      Here also we have studied a modified version of the Vicsek model from another point of view. In this version the interaction zone
   is quenched in space and that constitutes the only modification over the Vicsek model, all other 
   prescriptions of the Vicsek model remain unaltered. For example, for a two dimensional study, the entire space is divided 
   into a number of such interaction zones. While traveling, the trajectory of an agent passes through a series 
   of such interaction zones one after another. At an arbitrary instant of time, an agent interacts with all agents 
   within the interaction zone it is presently residing and similarly all agents within this zone interact among 
   themselves. Therefore, a common direction of motion is determined from these mutual interactions and it is 
   then assigned to all agents within this zone. At this point the noise appears into the picture and plays its 
   role. The direction of motion of each individual agent is then updated independently by applying the random 
   scalar noise. The quantitative description of the algorithm is as follows.

\begin{figure}[t]
\includegraphics[width=6.0cm]{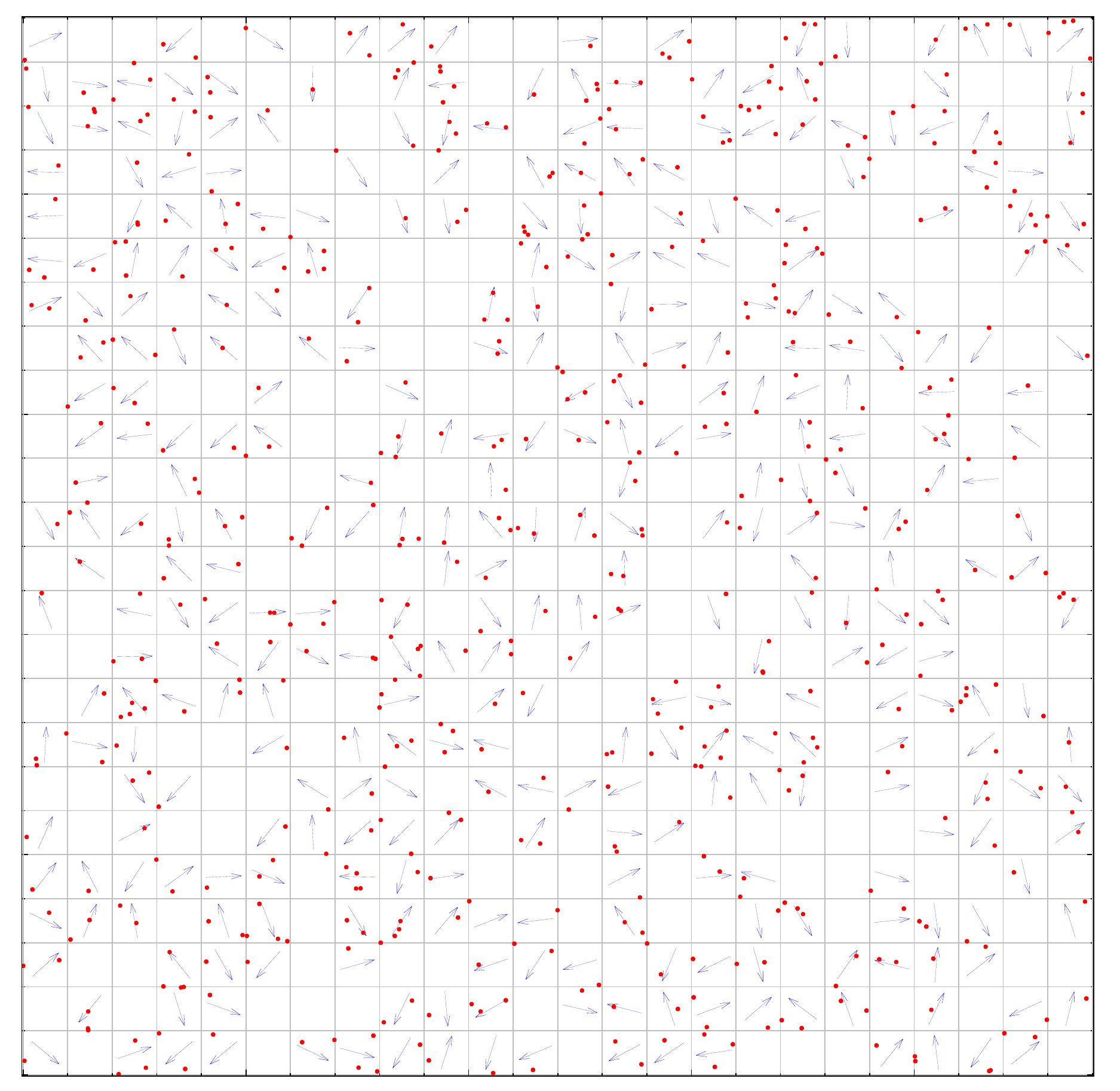}
\caption{A snapshot of the collection of agents at an arbitrary instant of time has been shown, marking them by the 
red dots, in a two dimensional system of size $L$ = 24 using the periodic boundary condition. The underlying space
has been discretized into small neighborhoods by the primitive cells of an $L \times L$ square lattice. For 
each cell the direction of the resultant of the velocity vectors of all agents has been shown by a blue arrow.
}
\label{fig01}
\end{figure}

      A collection of $N$ agents are released within a square box of size $L \times L$ on the $x-y$ plane at the positions 
   $(x_i,y_i),i=1,...,N$. The value of each co-ordinate is an independent and identically distributed random 
   number between 0 and $L$. The density $\rho = N/L^2$ of agents has been maintained to be unity in all 
   calculations in this paper. All agents have the same speed $v_0 = 1/2$ always and the orientation angles $\theta_i$ of 
   their velocity vectors have been assigned random values between 0 and $2\pi$ drawing them from a uniform probability
   distribution. A configuration of agents prepared in this way, constitute the initial state. The dynamical 
   state of the flock of agents is then evolved using a discrete time synchronous updating rule under periodic
   boundary condition, the time $t$ being the number of updates per agent. 

      At any arbitrary time $t$ an agent $i$ interacts with all $n_{\cal  R}(t)$ agents (including itself) 
   within a neighborhood ${\cal  R}$ around it. Unlike the Vicsek model here the neighborhoods are 
   quenched, i.e., they are fixed in space. We define these neighborhoods as the primitive cells of an underlying
   imaginary square lattice of size $L \times L$. More specifically, a typical neighborhood ${\cal  R}$ is the 
   primitive cell whose vertices are located at the coordinates $(x,y), (x+1,y), (x+1,y+1)$, and $(x,y+1)$ 
   where both $x$ and $y$ are the integer numbers. All $n_{\cal  R}(t)$ agents within a particular cell belong 
   to the same neighborhood and are neighbors of one another. 

\begin{figure}[t]
\includegraphics[width=6.0cm]{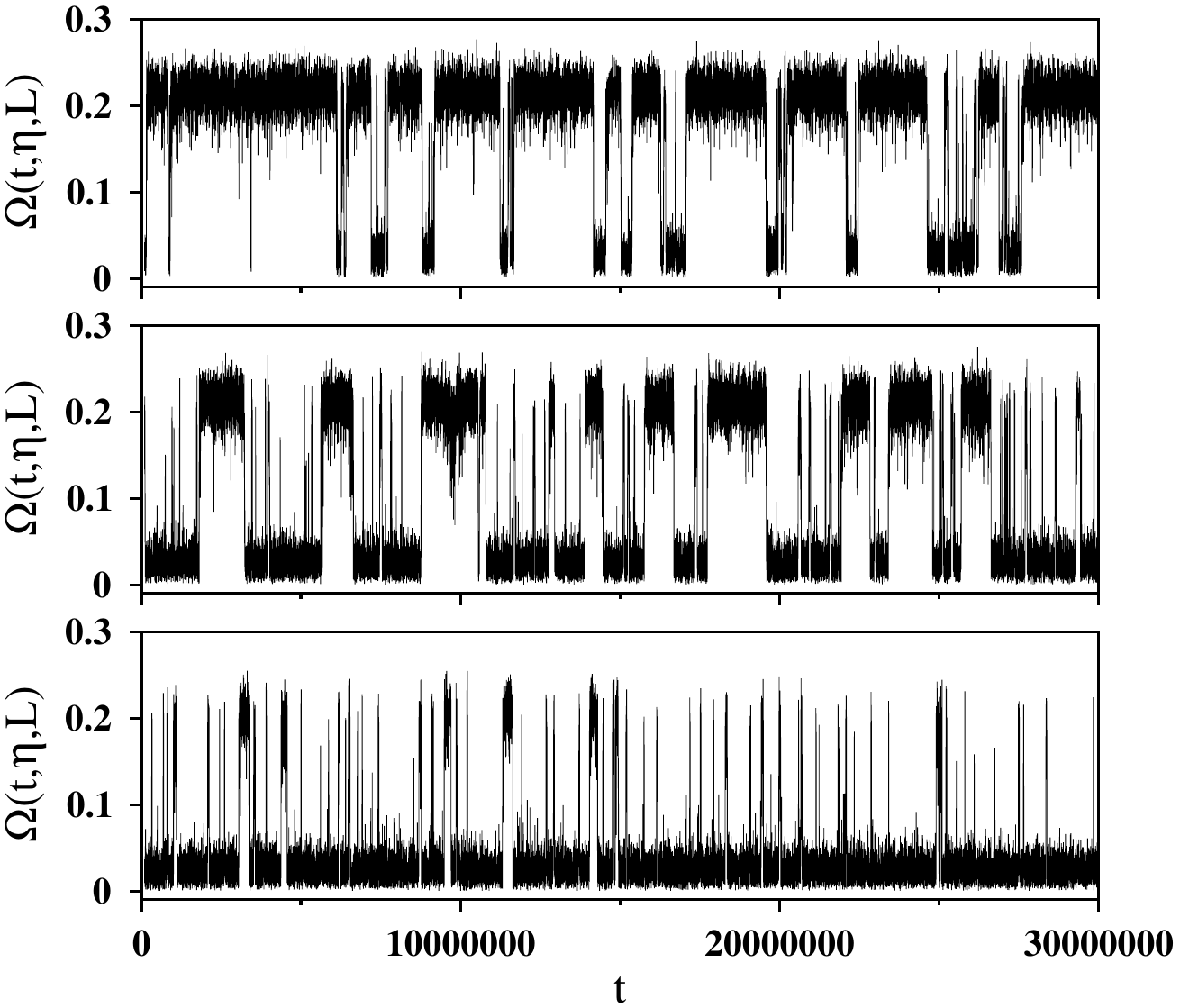}
\caption{$L = 128$: Variation of the instantaneous order parameter $\Omega(t,\eta,L)$ has been shown against time $t$
for the values of the noise strength $\eta =$ 2.262, 2.266 and 2.270 (from top to bottom). These $\eta$ values are very
close to $\eta_c = 2.266$. Here the system is seen to flip-flop between the ordered and disordered states corresponding
to non-zero and almost zero values of $\Omega$. These states are characterized by the presence and absence of high density 
correlated bands.
}
\label{fig02}
\end{figure}
   
      During the time evolution the system passes through a series of microstates defined by the specific positions 
   and the directions of motion of the $N$ agents. Let ${\bf v}_i(t)$ denote the velocity vector of the $i$-th agent at time $t$ which
   has the orientational angle $\theta_i(t)$. The orientational angles $\theta_i(t+1)$ at the next time step are then 
   estimated for all neighborhoods $\{{\cal R}\}$ in a synchronous manner. All agents $n_{\cal R}(t)$ within a 
   neighborhood mutually interact among themselves. The resultant of the velocity vectors of these agents is determined 
   and its orientational angle $\theta_i(t+1)$ is assigned to the directions of velocities of all agents (Fig. 1) as,
\begin {equation}
   \theta_i(t+1)=\tan^{-1}[\sum_{j\in {\cal R}} \sin \theta_j(t) / \sum_{j\in {\cal R}} \cos \theta_j(t)],
\end {equation}
   where the summation runs over all $n_{\cal R}(t)$ agents within ${\cal  R}$. Therefore, before the noise is switched on, all
   agents of a neighborhood ${\cal R}$ have the same velocity direction which is different in different neighborhoods.
   This should be compared with the original version of the Vicsek model where even before the application of noise, 
   different agents have different directions of motion since individual agents have distinctly different neighborhoods
   in general. This is the main difference between our quenched neighborhood version of the Vicsek model and its original 
   version. This modification has the numerical advantage since before the application of noise, the common direction of 
   motion of all agents within ${\cal R}$ is determined only once, which results the faster execution of the code. 

      However, on the introduction of scalar noise, the orientational angles become disordered. Along with the noise the 
   Eqn. (1) is modified as:
  \begin {equation}
\theta_i(t+1) = \tan^{-1}[\sum_{j\in {\cal R}} \sin \theta_j(t) / \sum_{j\in {\cal R}} \cos \theta_j(t)] + \zeta(\eta).
\end {equation} 
   The noise term $\zeta(\eta)$ quantifies the amount of error that is added to the orientational angle of each agent participating 
   in an interaction. Here $\eta$ measures the strength of the noise and $\zeta(\eta)$ represents a random angle for each agent drawn 
   from a uniform distribution within [-$\eta$/2, $\eta$/2]. Each agent is then displaced along its direction of motion 
   $\theta_i(t+1)$. In general, at every time instant, some agents leave a particular neighborhood ${\cal R}$ and 
   move to their adjacent neighborhoods. Similarly, another set of agents move into ${\cal R}$ from its adjacent 
   neighborhoods. 

\begin{figure}[t]
\includegraphics[width=6.0cm]{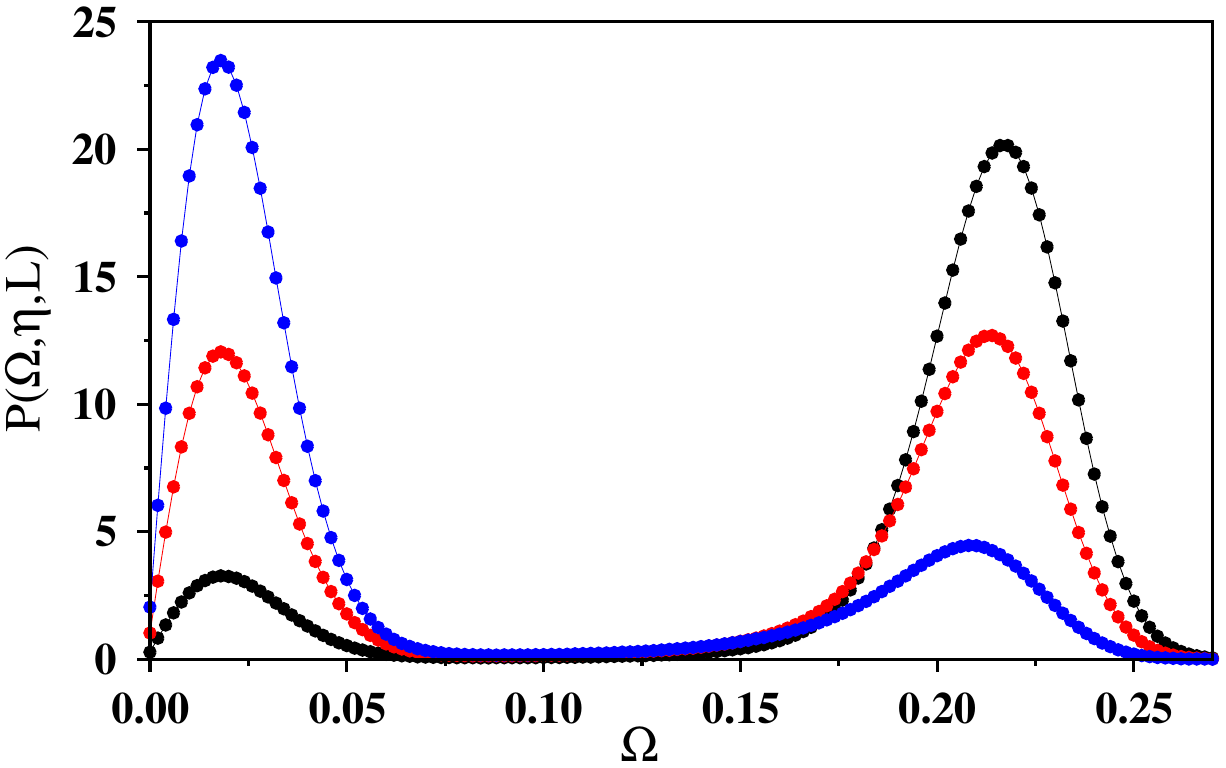}
\caption{$L = 128$: Probability distribution $P(\Omega,\eta,L)$ of the instantaneous order parameter in the stationary 
state against the order parameter $\Omega$. The three curves (from the right to the left) correspond to $\eta$ = 2.2610 
(black), 2.2645 (red), and 2.2680 (blue). Each curve has two peaks, one at a large $\Omega$ (ordered state) and the other 
at a small value of $\Omega$ (disordered state). 
}
\label{fig03}
\end{figure}

      The instantaneous global order parameter $\Omega(t, \eta, L)$ is defined for the entire system as the magnitude
   of the velocity vector of an agent, averaged over all agents and scaled by the speed $v_0$. 
\begin {align}
\Omega(t, \eta, L) & = \frac{1}{Nv_0} \bigg|\sum_{j\in N} {\bf v}_j(t)\bigg|.
\label{eqchilocal}
\end {align}
   In the stationary state $\Omega(t,\eta,L)$ is estimated over a long duration of time and is averaged to find 
   $\Omega(\eta,L)$.

\section {2. Description of the dynamical evolution}

      A computer code for the animation of the dynamical evolution of this system starting from the random initial state
   has been written and is run over long durations for different values of noise strengths. 
   Let us analyse the time evolution of the system for $\eta < \eta_c$. In particular, let us first consider
   the updating process of the directions of motion of the agents in two adjacent cells at locations $(x,y)$
   and $(x+1,y)$. Let the angles $\theta_i(t+1)$ for these cells before the application of the noise be
   approximately equal to $\pi/2$. Then after the application of noise and after moving one step, these two
   cells would exchange some agents. Since their directions are nearly the same, most of the agents of
   two adjacent cells would move to two new cells at $(x,y+1)$ and $(x+1,y+1)$ which are also adjacent cells.
   Thus we refer agents in two adjacent cells of nearly parallel directions of motion tend to stick together 
   maintaining their adjacency as the `cohesiveness property' of similarly moving agents in adjacent cells. 
   This cohesiveness is in-built in the dynamical rules of the collective motion. Because of this cohesiveness, 
   large clusters of agents gradually form as time passes. They move as a whole, and are extended spatially 
   across the direction of motion. Evidently, the most stable conformation of such a cluster appears when two 
   wings of it join together after wrapping the system because of the periodic boundary condition, which then 
   is called the `band'.

      In the absence of noise, the 
   agents move in the stationary state completely coherently in a ballistic fashion and $\Omega(0,L) = 1$. When the noise 
   is switched on and its strength $\eta$ is tuned to a small value, the motion of the individual agents in the stationary
   state is predominantly directed leading to a high value of the order parameter. In other words it means that on the 
   average the entire system of $N$ agents move along a globally fixed direction but the instantaneous velocity directions 
   of individual agents do fluctuate randomly with a small spread about this global 
   direction. On the other hand, when $\eta$ is tuned for the large values, individual agent's motion is grossly diffusive 
   and this leads to nearly vanishing values of the order parameter. The minimum value of noise parameter $\eta = \eta_c$ 
   where the order parameter vanishes for infinitely large system sizes, is called the critical point of the order-disorder 
   phase transition that takes place in this system of collective motion under the application of noise. 

\begin{figure*}[t]
\begin {tabular}{ccccc}
\includegraphics[width=3.0cm]{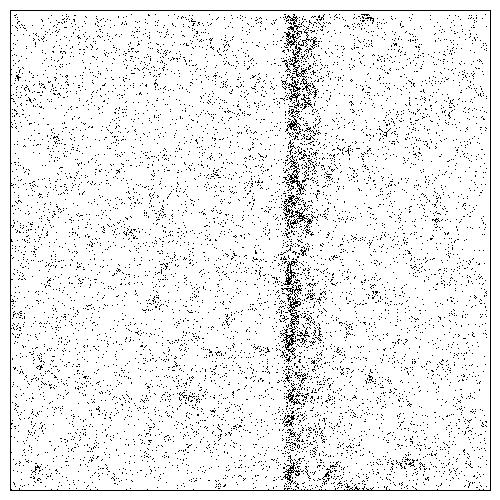} \hspace*{2.0mm}&
\includegraphics[width=3.0cm]{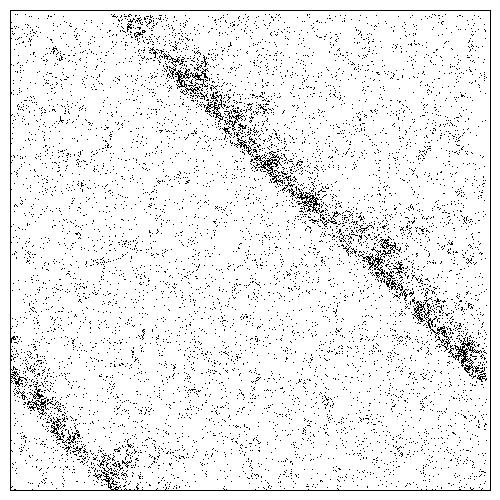} \hspace*{2.0mm}&
\includegraphics[width=3.0cm]{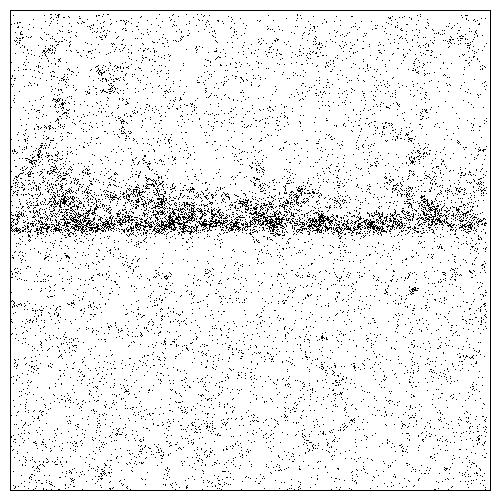} \hspace*{2.0mm}&
\includegraphics[width=3.0cm]{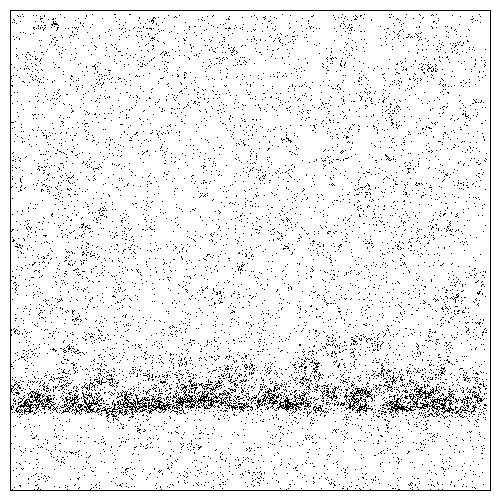} \hspace*{2.0mm}&
\includegraphics[width=3.0cm]{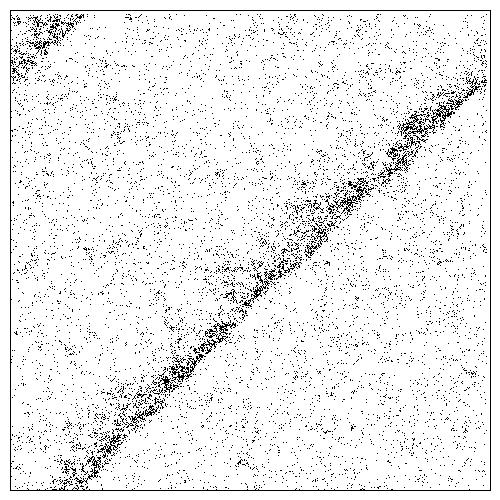} \hspace*{2.0mm}\\
2.262 & 2.140 & 2.100 & 2.080 & 2.060 \\
\includegraphics[width=3.0cm]{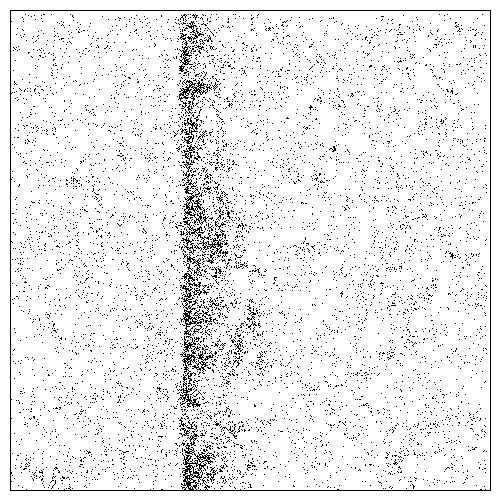} \hspace*{2.0mm}&
\includegraphics[width=3.0cm]{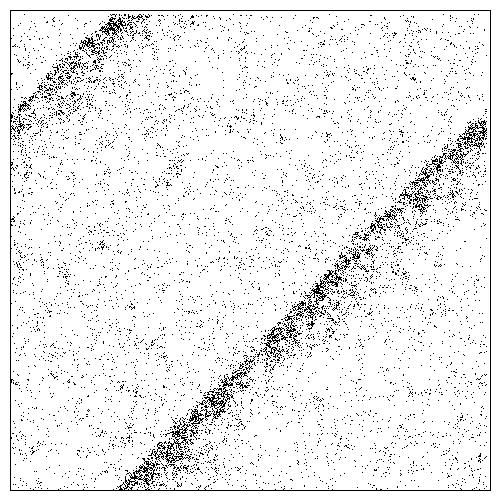} \hspace*{2.0mm}&
\includegraphics[width=3.0cm]{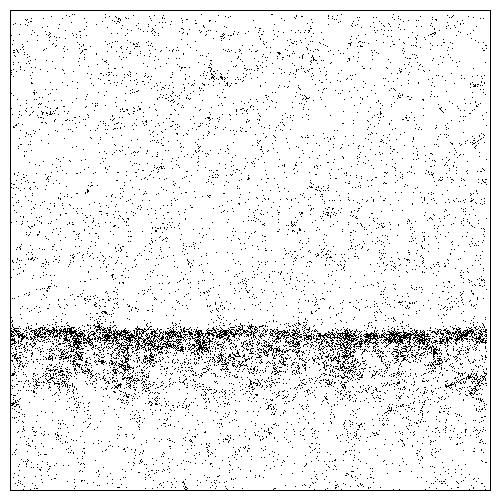} \hspace*{2.0mm}&
\includegraphics[width=3.0cm]{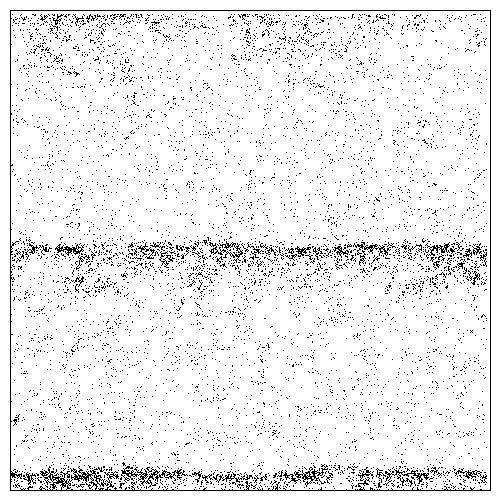} \hspace*{2.0mm}&
\includegraphics[width=3.0cm]{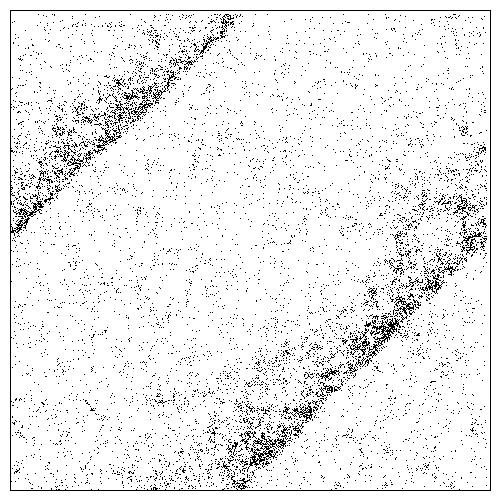} \hspace*{2.0mm}\\
2.020 & 1.950 & 1.900 & 1.850 & 1.800 \\
\includegraphics[width=3.0cm]{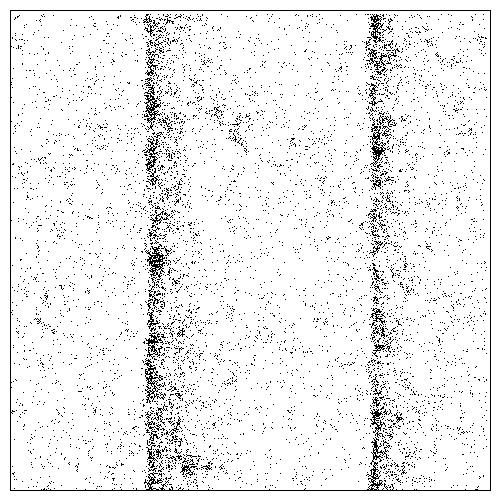} \hspace*{2.0mm}&
\includegraphics[width=3.0cm]{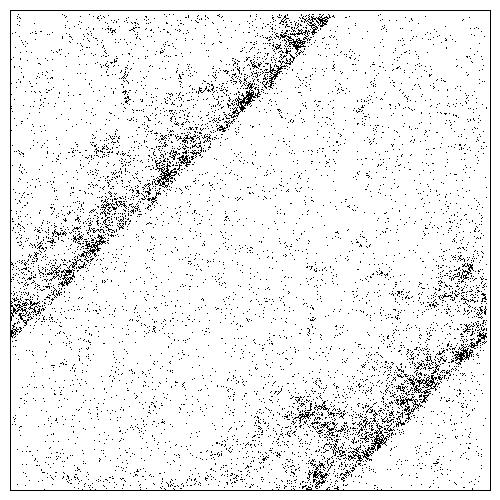} \hspace*{2.0mm}& 
\includegraphics[width=3.0cm]{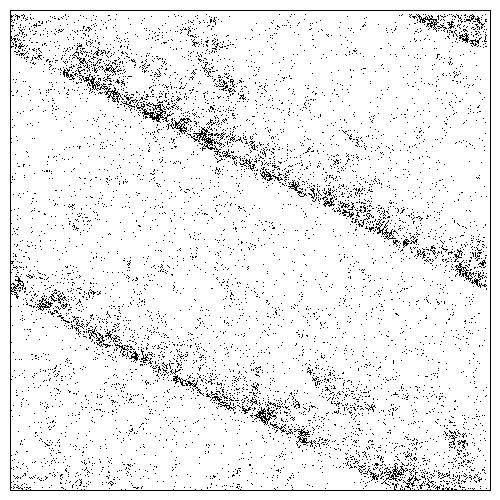} \hspace*{2.0mm}&
\includegraphics[width=3.0cm]{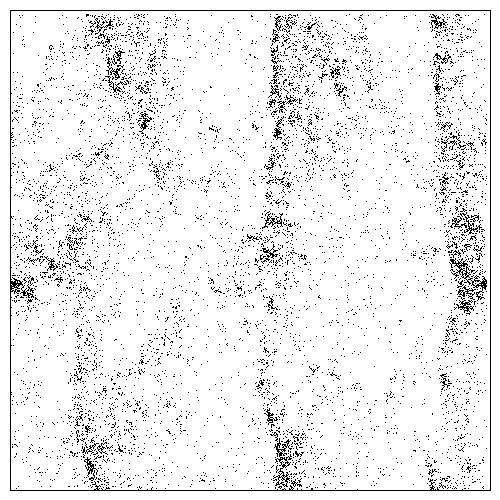} \hspace*{2.0mm}&
\includegraphics[width=3.0cm]{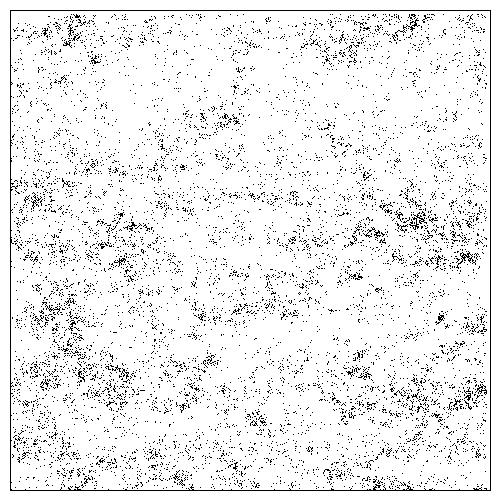} \hspace*{2.0mm}\\
1.740 & 1.700 & 1.600 & 1.250 & 0.500 \\
\end {tabular}
\caption{$L = 128$: Band structures have been shown for fifteen different stationary states evolved from the same 
initial state but subjected to noise strength $\eta$ of different magnitudes mentioned under the figure. Brief 
description of the individual bands are given in Table I.
}
\label{fig04}
\end{figure*}
\begin{table*}[t]
\begin{tabular}{lllll} \hline
$\eta$ & $\phi$    &$\Omega$    & Wrapping & Description\\ \hline 
2.262  & $\pi$     &  0.173     & $W(1,0)$ & Single vertical band.  \\
2.140  & $ 5\pi/4$ &  0.303     & $W(1,1)$ & Single diagonal band. \\
2.100  & $3\pi/2$  &  0.329     & $W(0,1)$ & Single horizontal band. \\
2.080  & $3\pi/2$  &  0.342     & $W(0,1)$ & Single horizontal band. \\
2.060  & $ 7\pi/4$ &  0.382     & $W(1,1)$ & Single diagonal band.\\
2.020  & $\pi$     &  0.379     & $W(1,0)$ & Single vertical band. \\
1.950  & $3\pi/4$  &  0.443     & $W(1,1)$ & Single diagonal band. \\
1.900  & $\pi/2$   &  0.450     & $W(0,1)$ & Single horizontal band. \\
1.850  & $\pi/2$   &  0.505     & $2W(0,1)$& Two parallel horizontal bands. \\
1.800  & $7\pi/4$  &  0.520     & $W(1,1)$ &Single diagonal band. \\
1.740  & $\pi$     &  0.562     & $2W(1,0)$&Two vertical bands. \\
1.700  & $7\pi/4$  &  0.568     & $W(1,1)$ &Single diagonal band. \\
1.600  & $3\pi/2 - arctan(1/2)$ & 0.617 & $W(1,2)$ &Single multiply wrapped band. \\
1.250  &           &  0.761     & & Parallel bands become blurred. \\
0.500  &           &  0.949     & &Bands are absent. \\ \hline
\end {tabular}
\caption{$L$ = 128: Brief descriptions of different types of bands.}
\label {TAB01}
\end{table*}

      How the system becomes increasingly ordered as the strength of the applied noise is systematically reduced?
   To understand it we need to follow the formation of bands. When $\eta$ is tuned to $\eta_c(L)$ a band of agents of high density 
   appears in the system for the first time. Within a band, the motion of the agents are directed. Such a band has the shape of a 
   thin and straight strip which moves as a whole along a specific direction perpendicular to the length of the band.
   Because of the imposed periodic boundary condition, the band takes the shape of a closed ring. Consequently, the 
   magnitude of the order parameter jumps discontinuously to a non-zero value when a band appears in the system. In 
   general, for $\eta < \eta_c$ the bands can be oriented along different directions, e.g., parallel to the sides, 
   along the diagonal directions, or even along some other directions. The shapes of the bands become increasingly 
   non-trivial as the system size becomes larger. For such large system sizes multiple bands can also be 
   simultaneously present. 

      A band consists of a set of directionally biased agents moving along almost (apart from noise) in the same direction. The entire 
   band moves in a sea of randomly diffusing agents. Thus, the whole $L \times L$ area is divided into two zones: the band, 
   comprised of the directed agents and the diffusive zone, comprising of the rest of the diffusive agents. 
   At every instant the band does two activities simultaneously: (i) It absorbs fresh agents into it at its front edge 
   picking them from the diffusive zone which execute a directed motion inside the band, and (ii) simultaneously it pushes the 
   directed agents at the back edge of the band into the diffusive zone. The rates of these two processes are equal 
   and in that way the width and shape of the band is maintained in the stationary state. The front edge of the band is
   sharp where as the back edge is hazy. Therefore, when there is only one band present in the system, an arbitrarily 
   tagged agent has two types of motion: for a certain amount of time it executes a diffusive motion, and then it is 
   swallowed by the band at the front edge. It then moves with the band for a little while as a directed agent,
   and then again dropped out in the diffusive region from the back edge. In the stationary state, this type of motion is repeated 
   ad infinitum for all agents.

\begin{figure}[t]
\includegraphics[width=8.5cm]{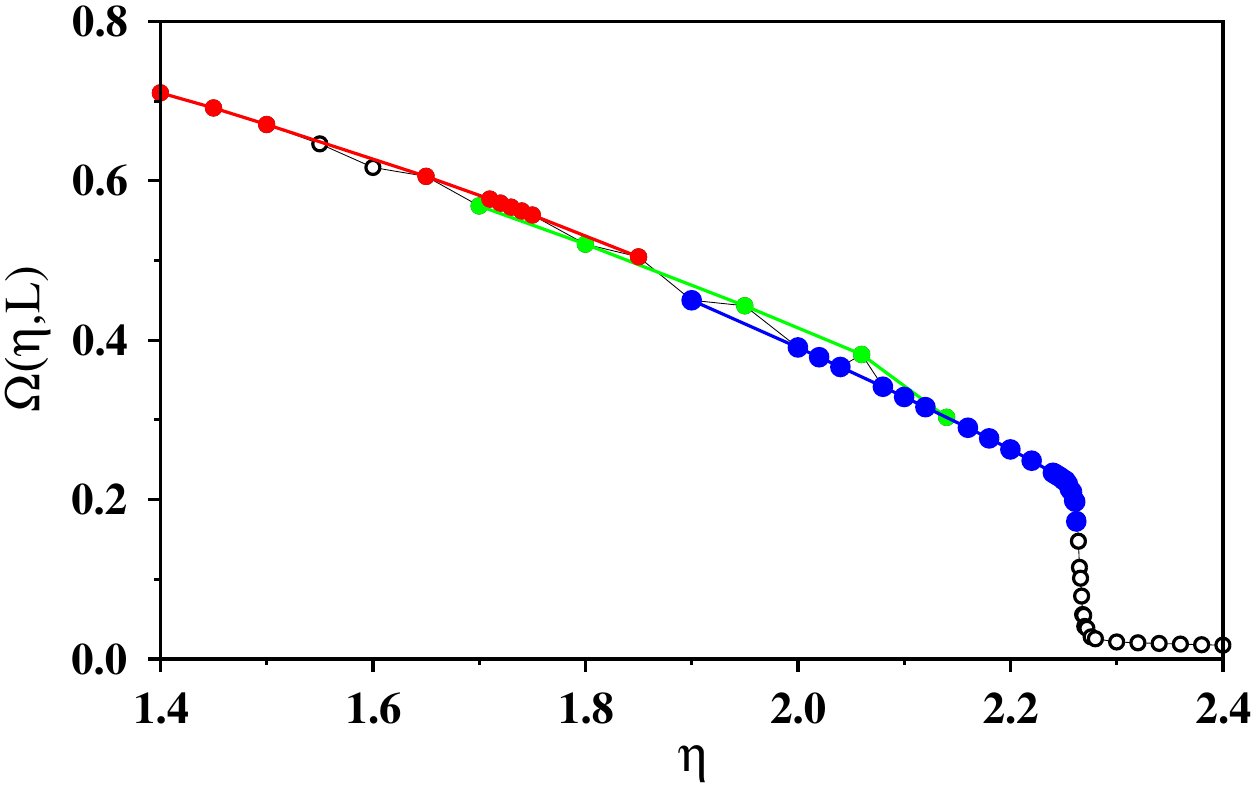}
\caption{$L = 128$: The average value of the order parameter $\Omega(\eta,L)$ in the stationary state has been plotted 
against the noise strength $\eta$ using black opaque circles. At the critical noise $\eta_c(L) = 2.2645$ there is a 
sharp rise in the order parameter. In the sub-critical regime three distinct data sets are identified which correspond 
to three different shaped bands and are plotted with circles filled with different colors: $W(1,0)$ and $W(0,1)$ (blue); 
$2W(1,0)$ and $2W(0,1)$ (red); $W(1,1)$ (green). In each set the $\Omega(\eta,L)$ increases almost linearly on 
decreasing $\eta$. 
}
\label{fig05}
\end{figure}
   
      In the following we report the results of our numerical study on three different system sizes, namely $L$ = 128, 256 
   and 512. We have observed how the discontinuous transition becomes more vivid and the band structure become increasingly rich as 
   the system size is systematically enlarged.

\section {3. Results}

\subsection {System size L = 128}

      We first exhibit the variation of the instantaneous order parameter $\Omega(t,\eta,L)$ against time $t$ in Fig. 2. 
   Three figures corresponding to three closely separated values of the noise parameter $\eta$ = 2.262, 2.266 and 2.270
   have been shown in the top, middle and bottom panels. The same random initial state has been used in all three cases. In each 
   case, the data has been plotted at the interval of every 1000 time steps and the time series has been shown for 30 
   million time steps. It is apparent from the plot that the system evolves through two possible metastable states, an 
   ordered state with high value of $\Omega$ and a disordered state with a very small value of $\Omega$. The system 
   flip-flops between these two states. It can also be observed that the system spends more time in the 
   ordered state with smaller noise at $\eta = 2.262$. On the other hand, the typical residence time in the disordered 
   state is longer with larger value of $\eta$ = 2.270. However, in between at $\eta$ = 2.266, the system resides in both 
   states almost equally frequently. Therefore, we approximately estimate $\eta_c$ = 2.266 as the critical noise parameter 
   of the system for $L = 128$.

\begin{figure}[t]
\includegraphics[width=6.0cm]{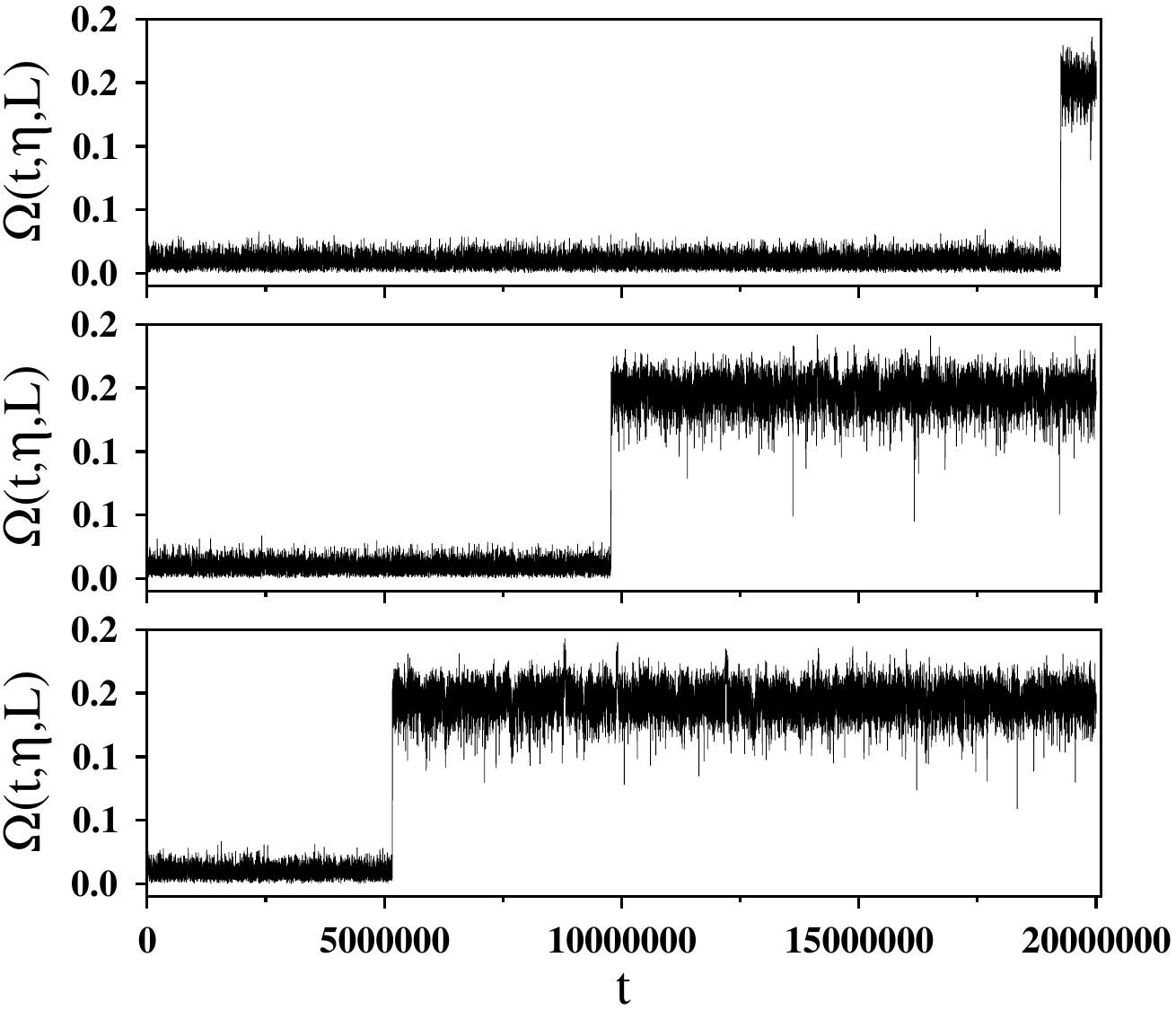}
\caption{$L = 256$: Plot of the instantaneous order parameter $\Omega(t,\eta,L)$ against time as in Fig. 2
for $\eta =$ 2.346, 2.348 and 2.350 (from top to bottom). Here the system is not seen to flip-flop between
two metastable states. Since the system evolves from a randomly selected state, initially the system is in 
a disordered state without any band having a nearly vanishing value of $\Omega$ but then it suddenly jumps 
to an ordered state on the appearance of a correlated band. 
}
\label{fig06}
\end{figure}

      To quantify the metastable states we have estimated the probability distribution of the order parameter 
   $P(\Omega,\eta,L)$ (Fig. 3). It has been found that for all three noise levels, the probability distribution has double 
   peaks at two distinct values of $\Omega$. These humps correspond to the ordered and disordered states. For 
   small $\eta = 2.2610$, the peak in the ordered state is taller than its peak in the disordered state. On the other hand, for 
   large $\eta = 2.2680$, it is the opposite, i.e., the peak in the disordered state is taller than its peak in the 
   ordered state. For the third plot with $\eta = 2.2645$ both peaks are approximately of same heights.

 \begin{figure*}[t]
 \begin {tabular}{ccccc}
 \includegraphics[width=3.0cm]{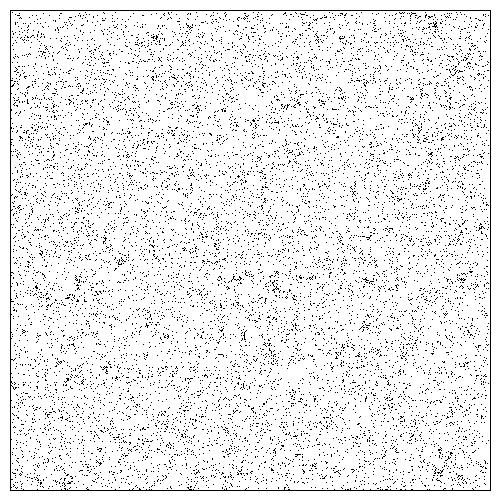} \hspace*{2.0mm}&
 \includegraphics[width=3.0cm]{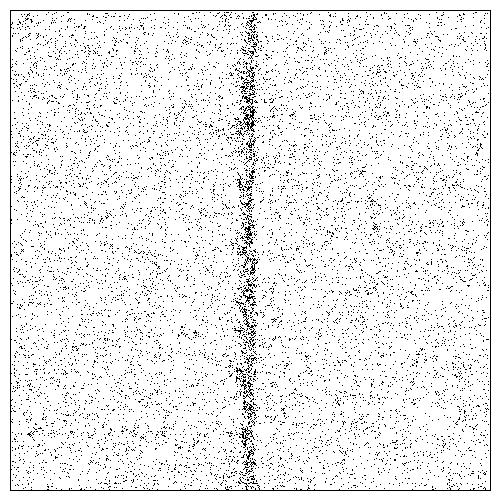} \hspace*{2.0mm}&
 \includegraphics[width=3.0cm]{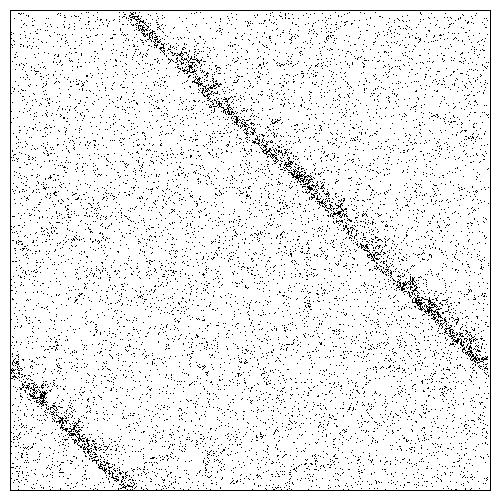} \hspace*{2.0mm}&
 \includegraphics[width=3.0cm]{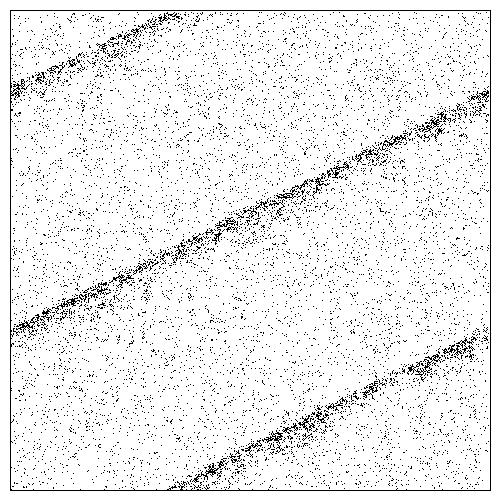} \hspace*{2.0mm}&
 \includegraphics[width=3.0cm]{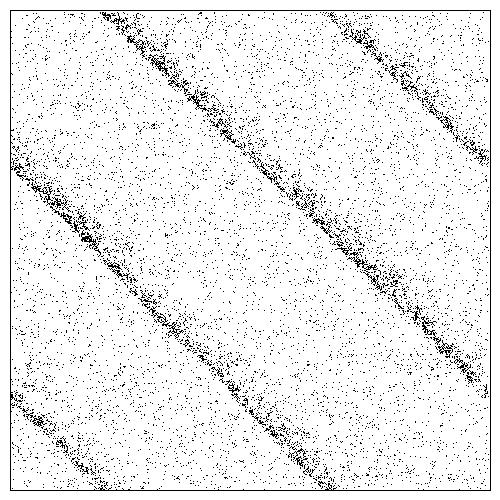} \hspace*{2.0mm}\\
 2.360 & 2.340 & 2.320 & 2.100 & 1.980 \\
 \includegraphics[width=3.0cm]{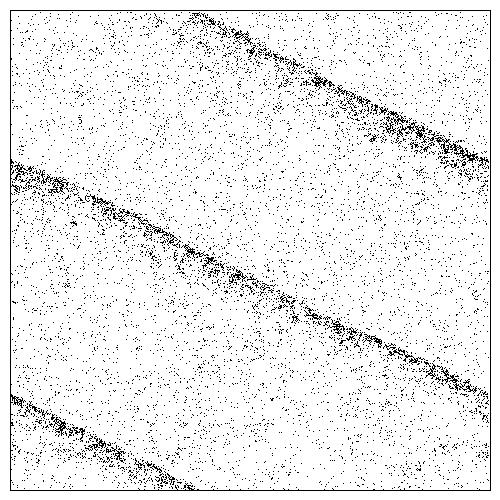} \hspace*{2.0mm}&
 \includegraphics[width=3.0cm]{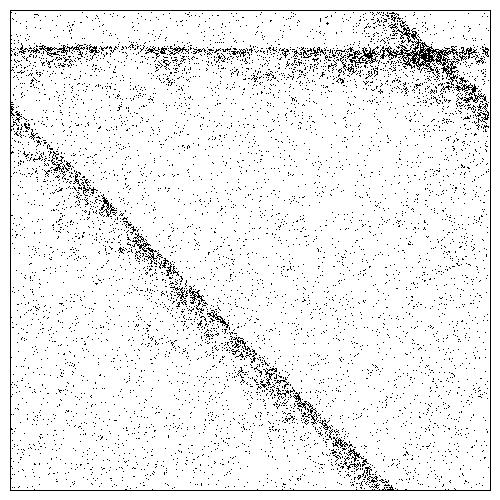} \hspace*{2.0mm}&
 \includegraphics[width=3.0cm]{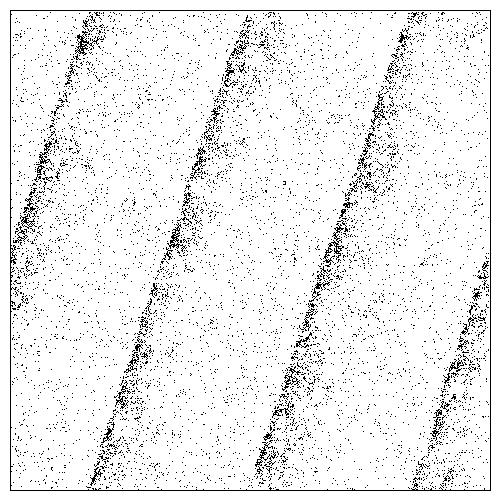} \hspace*{2.0mm}&
 \includegraphics[width=3.0cm]{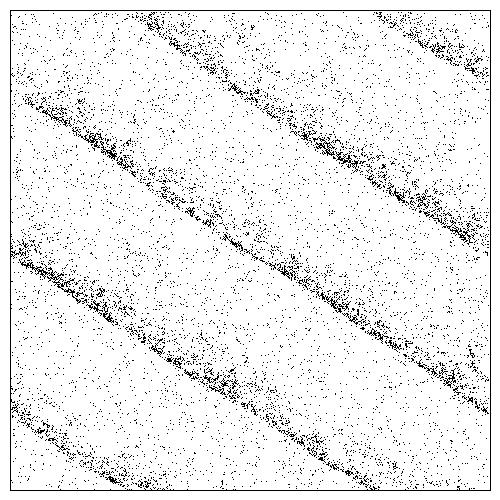} \hspace*{2.0mm}&
 \includegraphics[width=3.0cm]{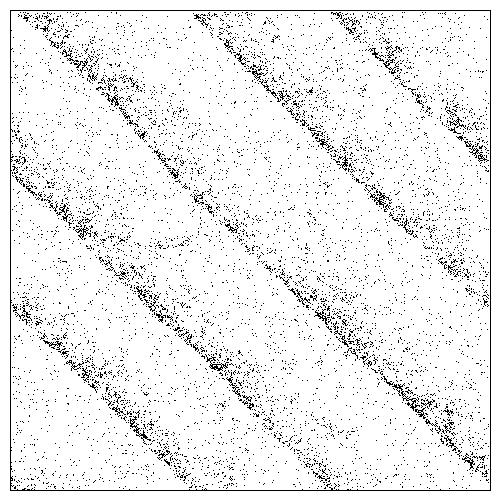} \hspace*{2.0mm}\\
 1.940 & 1.800 & 1.750 & 1.700 & 1.600 \\
 \includegraphics[width=3.0cm]{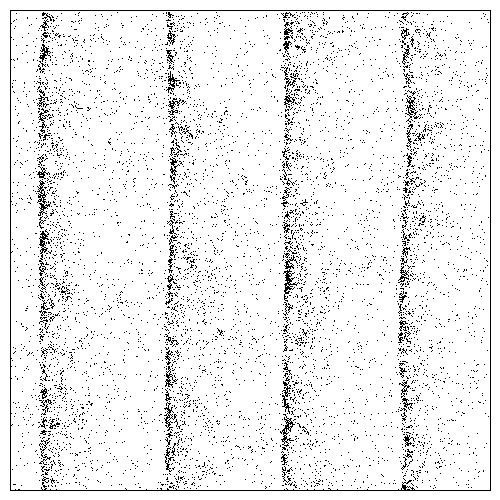} \hspace*{2.0mm}&
 \includegraphics[width=3.0cm]{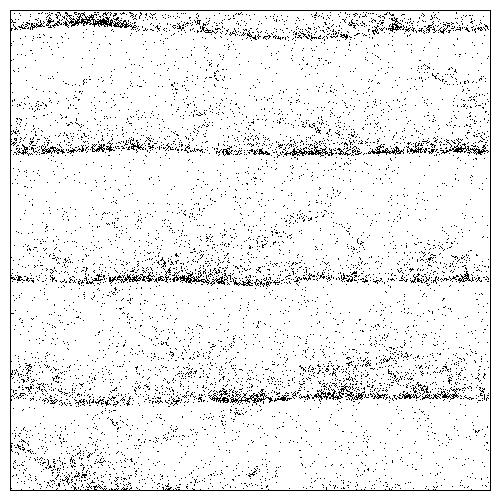} \hspace*{2.0mm}&
 \includegraphics[width=3.0cm]{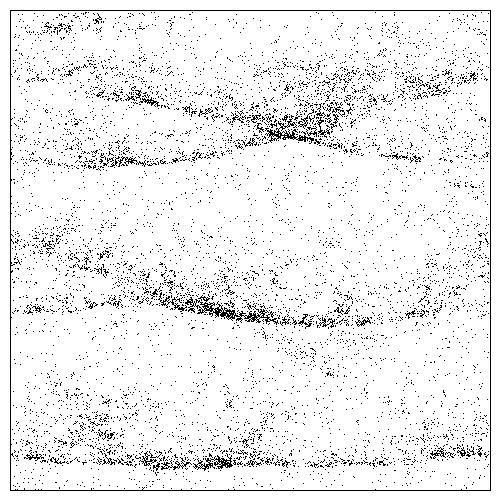} \hspace*{2.0mm}&
 \includegraphics[width=3.0cm]{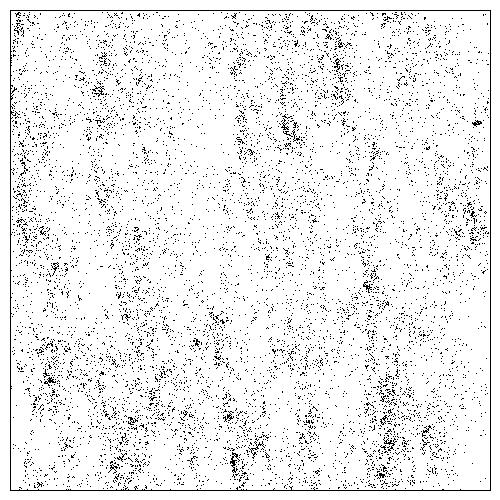} \hspace*{2.0mm}&
 \includegraphics[width=3.0cm]{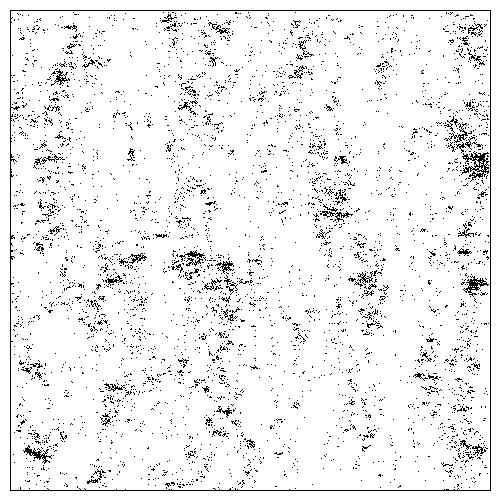} \hspace*{2.0mm}\\
 1.550 & 1.400 & 1.300 & 1.000 & 0.100 \\
 \end {tabular}
 \caption {$L$ = 256: Band structures in fifteen different stationary states evolved from the same initial positions and velocities
 but subjected to noise $\eta$ of different strengths have been exhibited using 16384 randomly selected agents. 
 The description of the 
 individual bands are given in Table II. 
 }
 \label{fig07}
 \end{figure*}
\begin{table*}[t]
\begin{tabular}{lllll} \hline
$\eta$ & $\phi$                 &  $\Omega$    & Wrapping  & Description \\ \hline 
2.360  &                        &  0.009       &           & Disordered state without band. \\
2.340  & 0                      &  0.149       & $W(1,0)$  & Single vertical bond.\\
2.320  & $5\pi/4$               &  0.189       & $W(1,1)$  & Single diagonal band. \\
2.100  & $\pi/2 + arctan(1/2)$  &  0.345       & $W(1,2)$  & Single multiply wrapped band. \\
1.980  & $5\pi/4$               &  0.429       & $2W(1,1)$ & Two parallel diagonal bands. \\
1.940  & $\pi/2 - arctan(1/2)$  &  0.437       & $W(1,2)$  & Single multiply wrapped band. \\
1.800  & $\pi/2,\pi/4$          &  0.488       & $W(0,1)$, $W(1,1)$ & Crossing: A diagonal and a horizontal band. \\
1.750  & $\pi - arctan(3)$      &  0.549       & $W(3,1)$  & Single multiply wrapped band. \\
1.700  & $3\pi/2 - arctan(2/3)$ &  0.577       & $W(2,3)$  & Single band, wrapped four times. \\
1.600  & $5\pi/4$               &  0.620       & $3W(1,1)$ & Three parallel diagonal bands. \\
1.550  & $\pi$                  &  0.647       & $4W(1,0)$ & Four parallel vertical bands. \\
1.400  & $3\pi/2$               &  0.710       & $4W(0,1)$ & Four parallel horizontal bands. \\
1.300  &                        &  0.739       &           & Some distorted and hazy bands. \\
1.000  &                        &  0.838       &           & Ordered state without band. \\
0.100  &                        &  0.997       &           & Ordered state without band. \\ \hline
\end {tabular}
\caption{$L$ = 256: Brief descriptions of different types of bands.}
\label {TAB02}
\end{table*}

      In Fig. 4 snapshots of fifteen agent configurations have been shown, and the characterization of each figure has been done in
   Table 1. Presence of bands have been searched for in the stationary states of the system. We started from a high value
   of $\eta$ and the reduced its value systematically in small intervals. The first high density stable band is observed for 
   $\eta=2.262$. 
   The flip-flop dynamics of the system exhibited in Fig. 2 implies the appearance and disappearance of such bands with time. 
   The orientation of the band is measured by the angle $\phi$ between the direction of motion of the band and the positive $x$ 
   direction. All bands always move along the normal to the front edge of the band. Therefore, for the first few snapshots of 
   Fig. 4, the angle $\phi$ has the values $\pi$, $5\pi/4$, $3\pi/2$, $3\pi/2$, $7\pi/4$, ... etc. The first diagonal band 
   appeared at $\eta = 2.140$. The first parallel double bands appeared at $\eta = 1.850$. The next non-trivial band
   appeared at $\eta = 1.600$ with $\phi = 3\pi/2 - arctan(1/2)$.

      The bands are stable and with the periodic boundary condition imposed, they wrap the system different number of times in 
   different snapshots. For example, if the orientation of the band is neither horizontal nor vertical, it must be oriented at 
   an angle $\phi$ such that it wraps the system. We characterize such a wrapped band by an integer pair $W(m,n)$ such that the 
   numbers of its intersection are $m$ and $n$ with the $x$ and $y$ axes respectively and $\phi$ differs from arctan($m/n$) 
   either by $\pm \pi/2$ or by $\pm \pi$. For example, $W(1,0)$ and $W(0,1)$ are the vertical and horizontal bands respectively. 
   A single diagonal band is denoted by $W(1,1)$. A set of $k$ parallel $W(m,n)$ bands are denoted by $kW(m,n)$.

\begin{figure}[t]
\includegraphics[width=8.5cm]{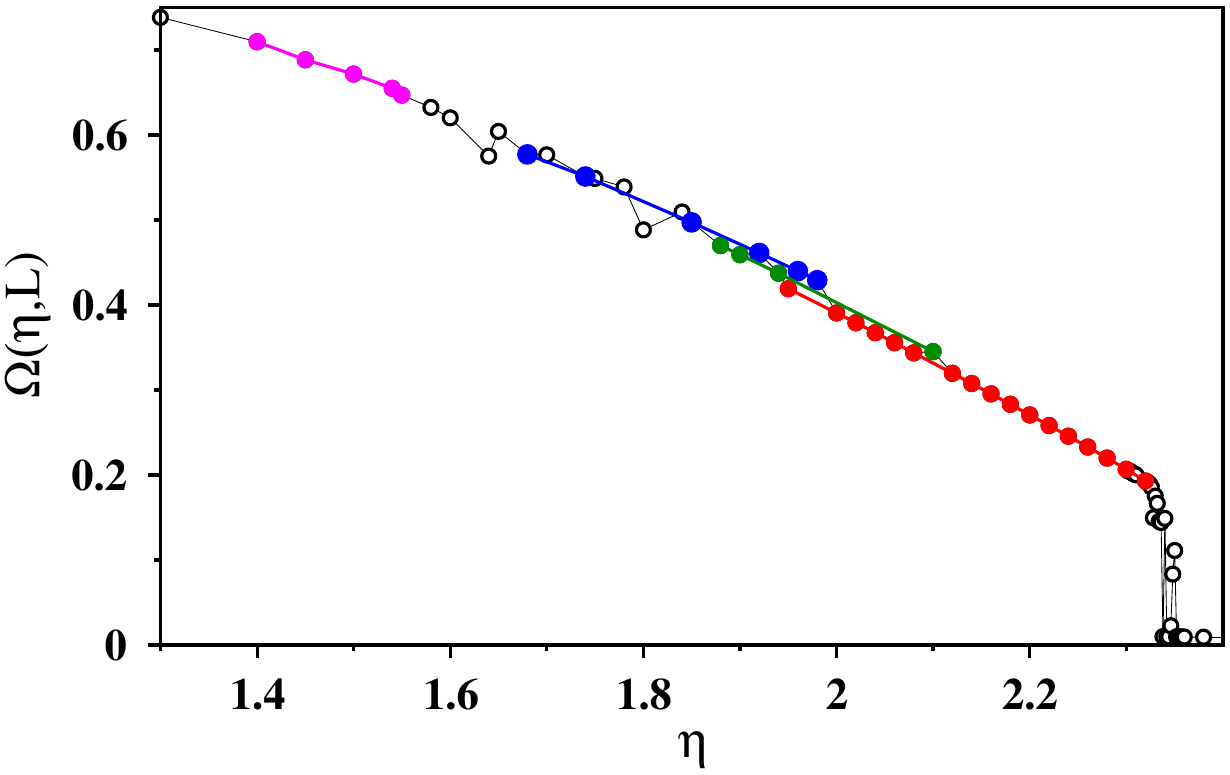}
\caption{$L = 256$: The average value of the order parameter $\Omega(\eta,L)$ in the stationary state has been plotted 
against the noise strength $\eta$ using black opaque circles. At the critical noise $\eta_c(L) = 2.350$ there is a 
sharp rise in the order parameter similar to a discontinuous transition. In the subcritical regime four different data 
sets are identified with four distinctly different shaped bands and are represented by filled circles of different colors:
$W(1,1)$ (red); $2W(1,1)$ (blue); $4W(1,0)$ and $4W(0,1)$ (magenta) and $W(1,2)$ and $W(2,1)$ (green). In each set the 
$\Omega(\eta,L)$ increases almost linearly on decreasing $\eta$. 
}
\label{fig08}
\end{figure}

      Therefore, it is apparent that as $\eta$ decreases the agents become more strongly correlated. Such stronger correlation
   appears in longer lengths as well as wider widths of the bands. Long band lengths are accommodated by increasing the number 
   of bands, selecting the non-trivial orientation of the bands, or by increasing the wrapping numbers. When $\eta$ is tuned 
   down to $\approx$ 1.250, the distinct structure of bands start vanishing, i.e., the dismantling process of the bands start. 
   Evidently, no distinct band was observed when $\eta$ was set to even smaller values.

      The variation of the order parameter $\Omega(\eta,L)$ for $L = 128$ against the noise strength $\eta$ has been shown 
   in Fig. 5. For $\eta < \eta_c(L)$ the order parameter $\Omega(\eta,L)$ increases almost linearly on reducing $\eta$. 
   It is clear that the entire plot is the combination of different subset of points corresponding to different shaped
   bands. In Fig. 5 we have marked three sets of colored circles that represent the data for three types of bands. 
   For $\eta > \eta_c(L)$ the $\Omega(\eta,L)$ assumes a nearly constant value close to zero on increasing $\eta$.

\begin{figure}[t]
\includegraphics[width=6.0cm]{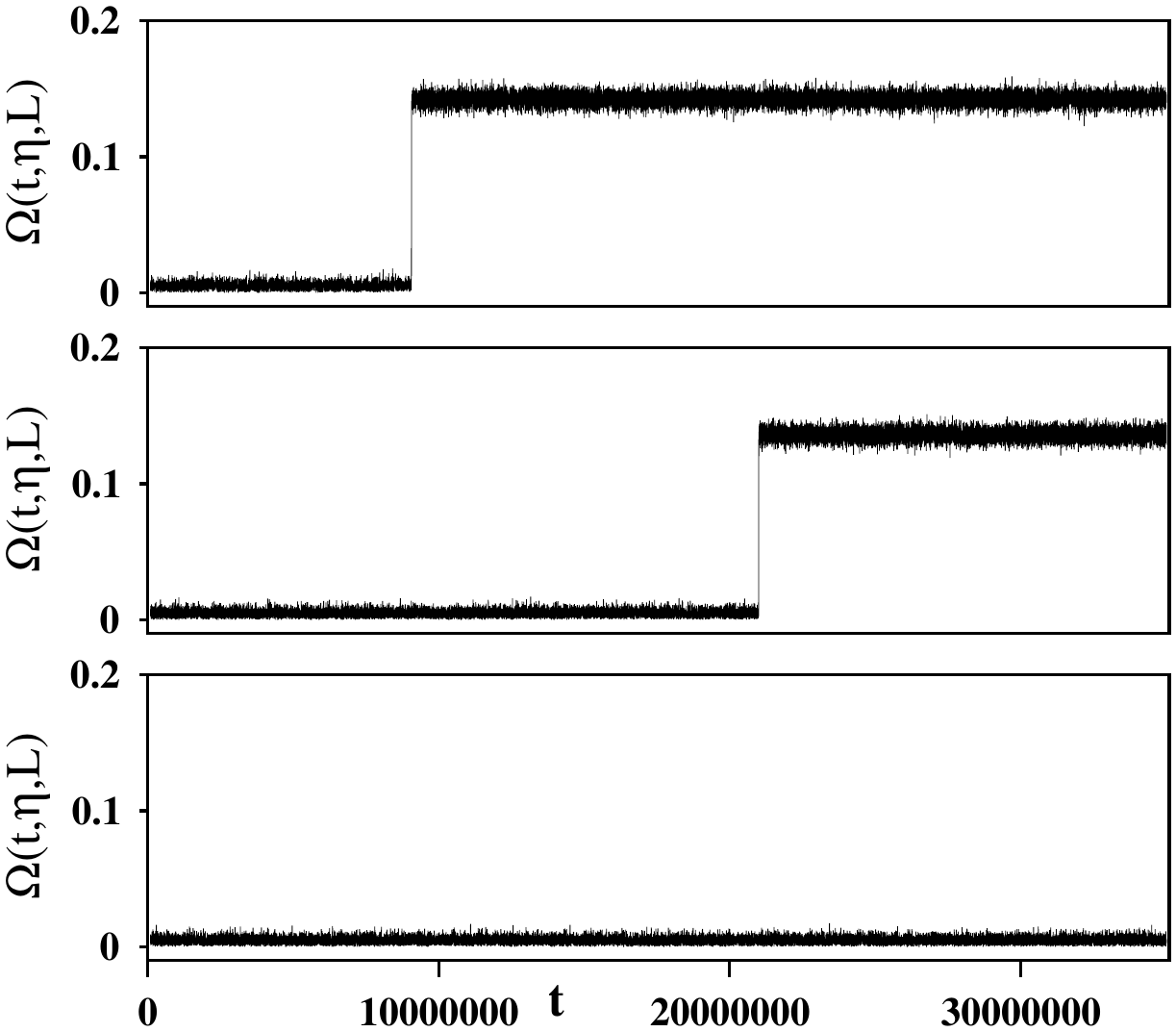}
\caption{$L = 512$: Variation of the instantaneous order parameter $\Omega(t,\eta,L)$ has been shown against time $t$
and for $\eta =$ 2.350, 2.360, and 2.370 (from top to bottom). As $\eta$ approaches $\eta_c$, increasingly longer time is 
required for switching over from the disordered to the ordered state.
}
\label{fig09}
\end{figure}

\begin{figure*}[t]
\begin {tabular}{ccccc}
\includegraphics[width=3.0cm]{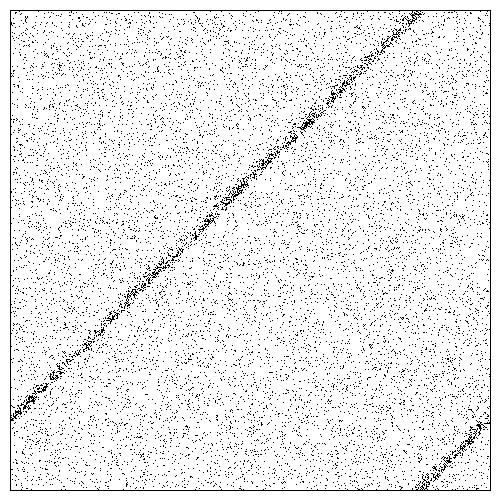} \hspace*{2.0mm}& 
\includegraphics[width=3.0cm]{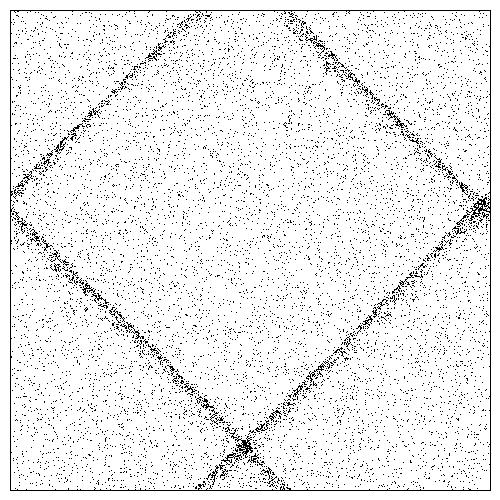} \hspace*{2.0mm}&
\includegraphics[width=3.0cm]{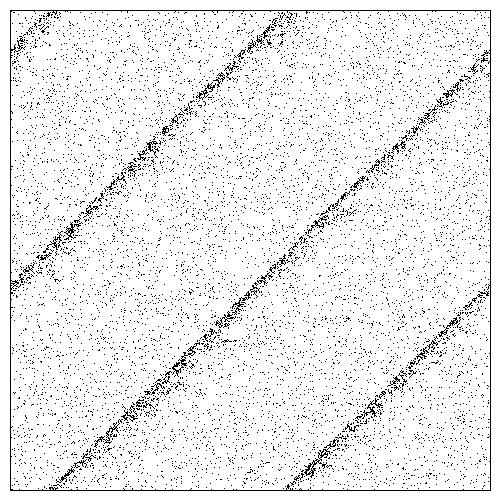} \hspace*{2.0mm}&
\includegraphics[width=3.0cm]{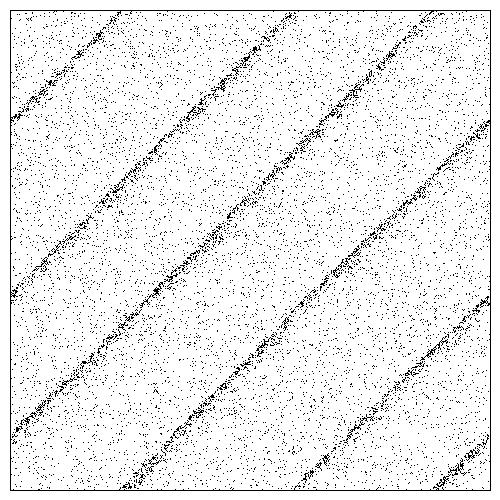} \hspace*{2.0mm}& 
\includegraphics[width=3.0cm]{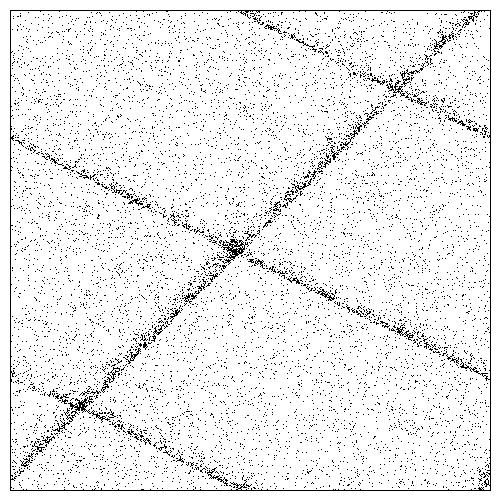} \hspace*{2.0mm}\\
2.368 & 2.130 & 2.110 & 2.100 & 2.090 \\
\includegraphics[width=3.0cm]{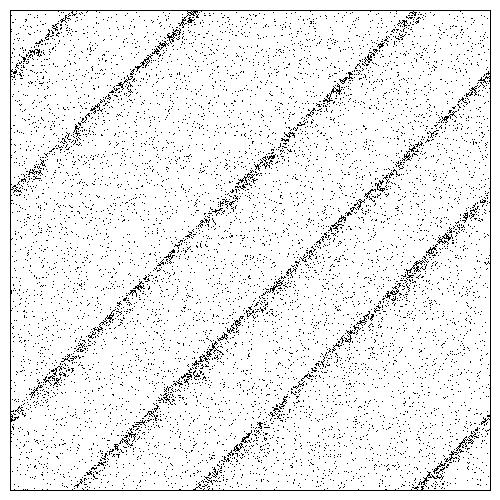} \hspace*{2.0mm}& 
\includegraphics[width=3.0cm]{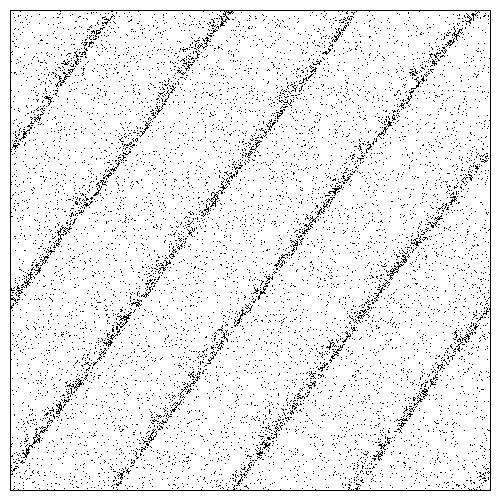} \hspace*{2.0mm}&
\includegraphics[width=3.0cm]{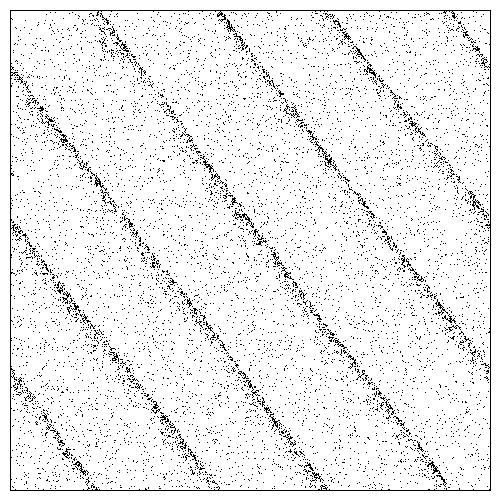} \hspace*{2.0mm}& 
\includegraphics[width=3.0cm]{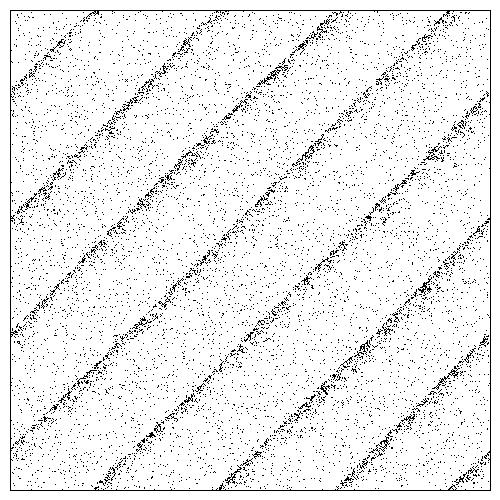} \hspace*{2.0mm}&
\includegraphics[width=3.0cm]{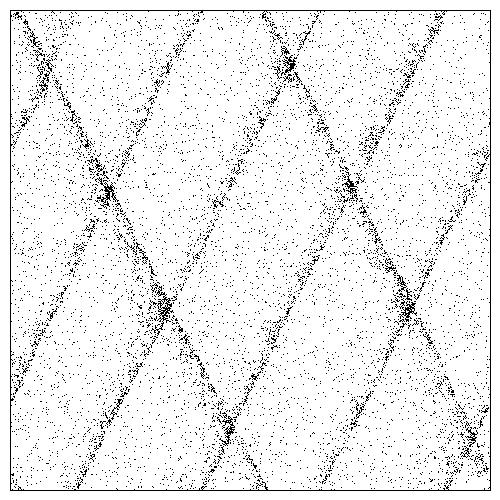} \hspace*{2.0mm} \\
2.040 & 2.000 & 1.980 & 1.900 & 1.800 \\
\includegraphics[width=3.0cm]{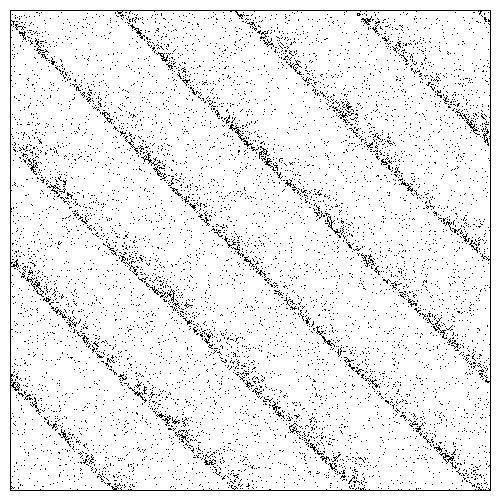} \hspace*{2.0mm}& 
\includegraphics[width=3.0cm]{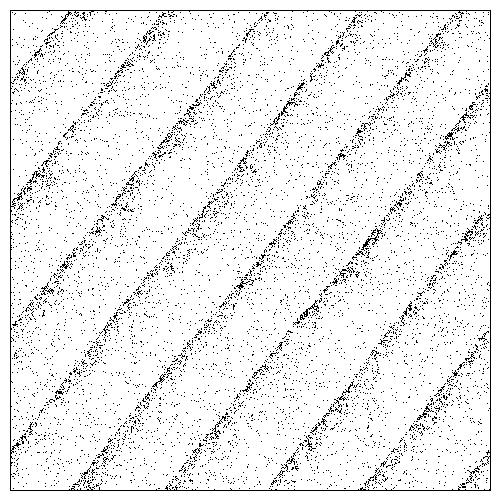} \hspace*{2.0mm}&
\includegraphics[width=3.0cm]{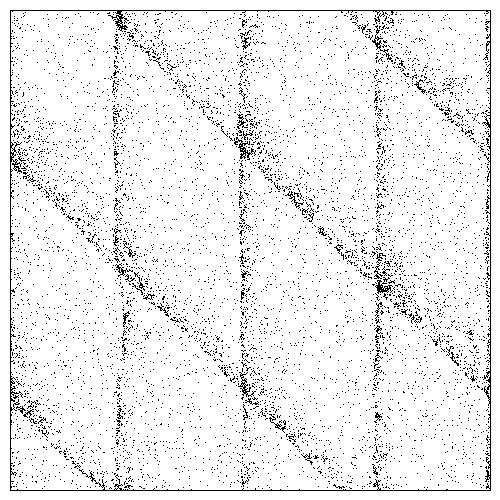} \hspace*{2.0mm}& 
\includegraphics[width=3.0cm]{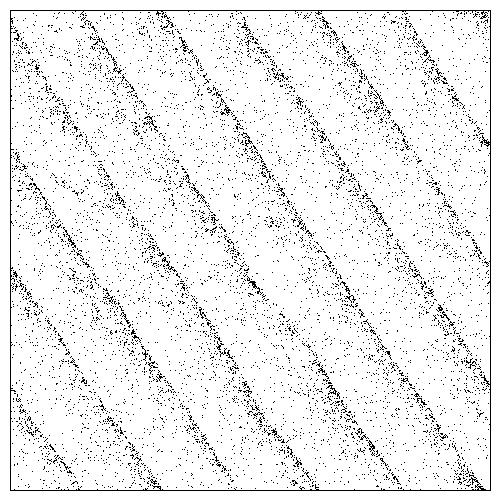} \hspace*{2.0mm}&
\includegraphics[width=3.0cm]{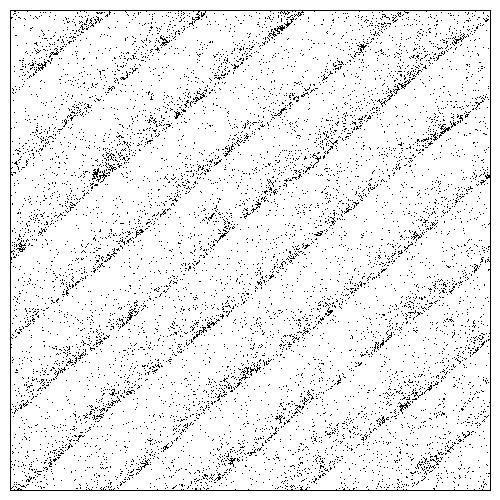} \hspace*{2.0mm} \\
1.750 & 1.700 & 1.650 & 1.550 & 1.500 \\
\includegraphics[width=3.0cm]{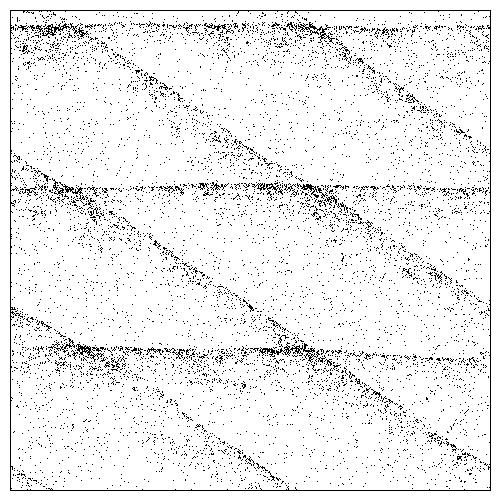} \hspace*{2.0mm}& 
\includegraphics[width=3.0cm]{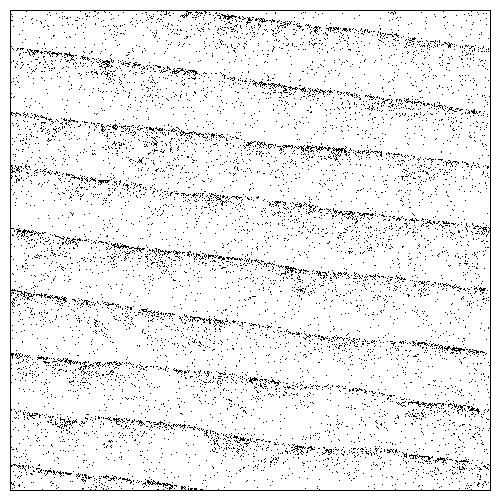} \hspace*{2.0mm}&
\includegraphics[width=3.0cm]{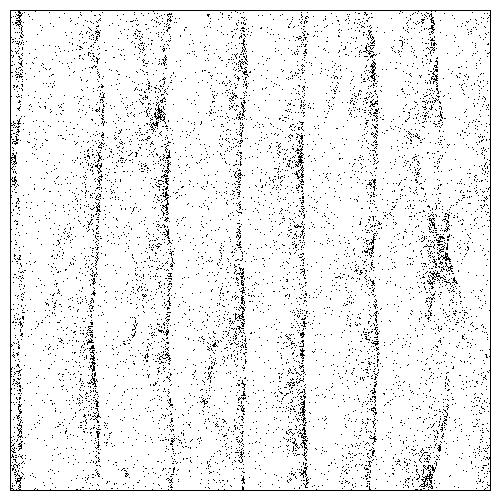} \hspace*{2.0mm}& 
\includegraphics[width=3.0cm]{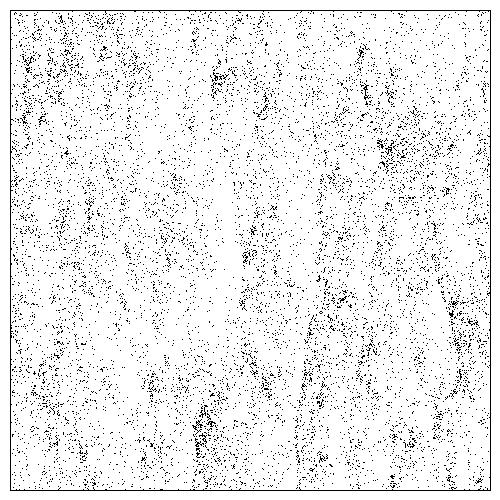} \hspace*{2.0mm}&
\includegraphics[width=3.0cm]{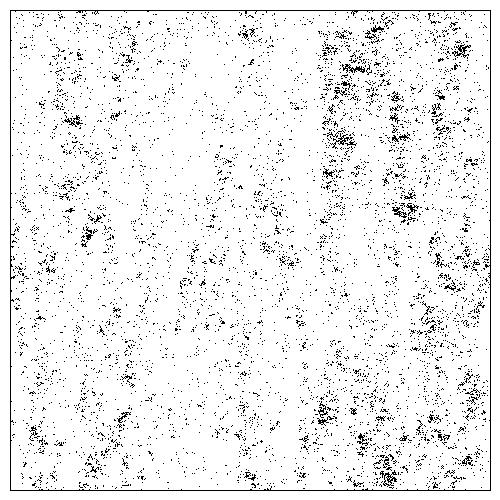} \hspace*{2.0mm} \\
1.450 & 1.400 & 1.350 & 1.000 & 0.100 \\
\end {tabular}
\caption{$L$ = 512: Band structures in twenty different stationary states evolved from the same initial positions and velocities
but subjected to noise $\eta$ of different strengths have been exhibited using 16384 randomly selected agents. The description of the
individual bands are given in Table III.
}
\label{fig10}
\end{figure*}
\begin{table*}[t]
\begin{tabular}{lllll} \hline
$\eta$  & $\phi$                       & $\Omega$  & Wrapping     &Description \\ \hline 
2.368   & $7\pi/4$                     & 0.115     &W(1,1)        & Single diagonal band. \\
2.130   & $\pi/4,4\pi/4$               & 0.227     &W(1,1),W(1,1) & Crossing of two diagonal bands at right angle. \\
2.110   & $3\pi/4      $               & 0.326     &2W(1,1)       & Two parallel diagonal bands.\\
2.100   & $\pi/4$                      & 0.350     &3W(1,1)       & Three parallel diagonal bands.\\
2.090   & $7\pi/4,3\pi/2-arctan(1/2)$  & 0.288     &W(1,1),W(1,2) &  Crossing of two sets of bands. \\
2.040   & $3\pi/4$                     & 0.384     &3W(1,1)       & Three parallel diagonal bands. \\
2.000   & $3\pi/2 + arctan(4/3)$       & 0.413     &W(4,3)        & A single band that wraps multiple times.  \\
1.980   & $\pi/2 - arctan(4/3)$        & 0.424     &W(4,3)        & A single band that wraps multiple times.  \\
1.900   & $3\pi/4$                     & 0.472     &4W(1,1)       & Four parallel diagonal bands.  \\
1.800   & $\pi/2 - arctan(2),3\pi/2+arctan(2)$ &0.489&W(2,1) and 2W(2,1)& Crossing of two sets of bands.\\ 
1.750   & $5\pi/4$                     & 0.546     &4W(1,1)       & Four parallel diagonal bands.\\
1.700   & $\pi - arctan(5/4) $         & 0.574     &W(5,4)        & A single band that wraps multiple times.  \\
1.650   & $\pi, 5\pi/4$                & 0.570     &4W(1,0),2W(1,1)& Crossing of two sets of bands. \\
1.550   & $\pi/2 - arctan(6/4)$        & 0.642     &2W(3,2)       & Two parallel multiply wrapped bands.\\
1.500   & $3\pi/2 + arctan(5/6)$       & 0.663     &W(5,6)        & A single band that wraps multiple times. \\
1.450   & $\pi/2, \pi/2 - arctan(3/2)$ & 0.670     &3W(0,1),W(2,3)& Crossing of two sets of bands.\\
1.400   & $\pi/2 - arctan(1/8)$        & 0.709     &W(1,8)        & Single multiply wrapped band.\\
1.350   &                              & 0.724     &7W(1,0)       & Seven parallel vertical bands.\\
1.000   &                              & 0.838     &              & No band structure is found.\\
0.100   &                              & 0.996     &              & No band structure is found.\\ \hline
\end {tabular}
\caption{$L$ = 512: Brief description of different types of bands.}
\label {TAB03}
\end{table*}
\begin{figure}[t]
\includegraphics[width=8.5cm]{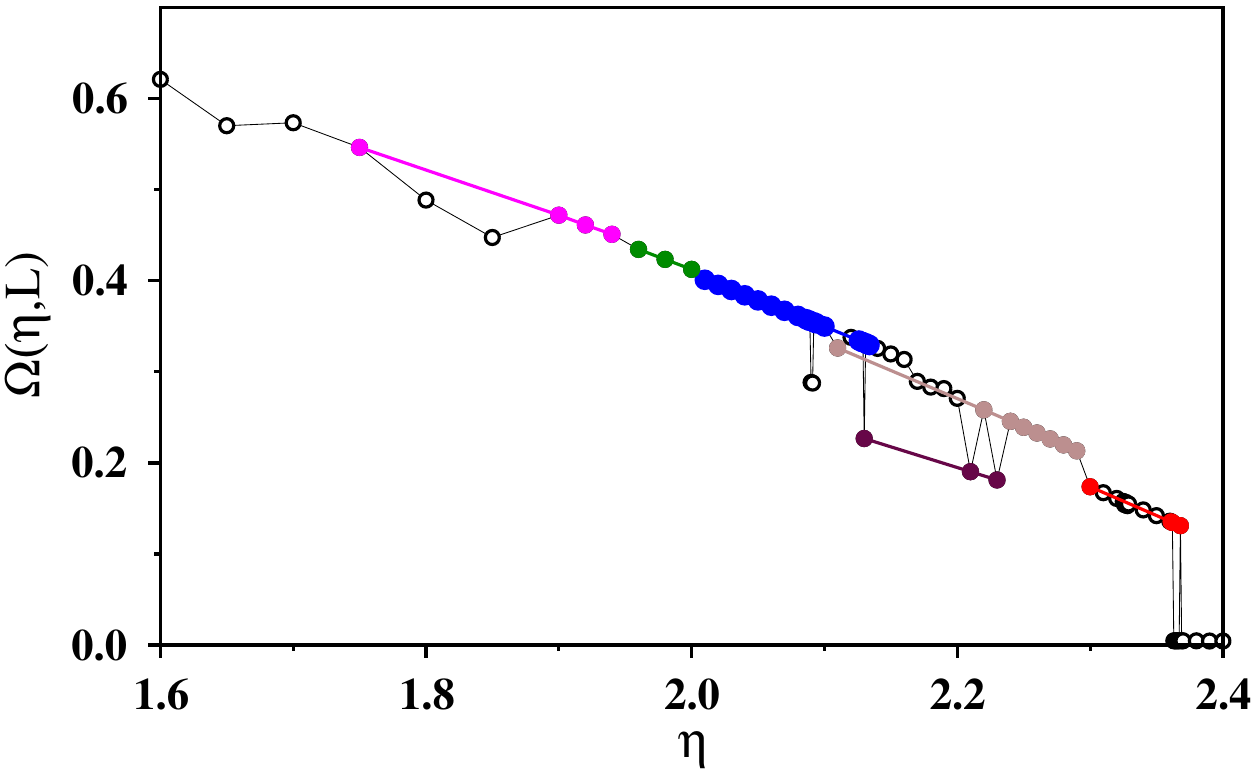}
\caption{$L$ = 512: The stationary state order parameter $\Omega(\eta,L)$ has been plotted against the noise strength 
$\eta$. The critical value $\eta_c(L)$ of transition has been found to 2.368. In the sub-critical regime six distinct 
sub-sets of data points have been identified for six differently shaped bands. They are grouped together and labeled
by their wrapping numbers: $W(1,1)$ (red); $2W(1,1)$ (brown); $3W(1,1)$ (blue); $4W(1,1)$ (magenta); $W(4,3), W(3,4)$ 
(green) and $W(1,1), W(1,1)$ bands are presented by maroon color.
}
\label{fig11}
\end{figure}

\subsection {System size L = 256}

      Here also the dynamics starts from the same completely random initial state for all values of the noise strength parameter $\eta$, 
   and therefore the motion of the agents are predominantly diffusive at the early stage. As time is elapsed, the system takes time to 
   organize itself. Typically, after a substantial amount of relaxation time, bands are formed here as well. In contrast to the 
   situation in $L = 128$ system, here the system does not flip-flop between two metastable states. Consequently, the magnitude of the 
   order parameter jumps up only for once from a nearly vanishing value to a finite magnitude. Three such jumps have been shown in Fig. 
   6 where the time series for the order parameter has been plotted against time for the 20 million time steps. For the noise level of 
   $\eta$ = 2.346, 2.348, and 2.350 the transitions take place at times $\approx$ 18, 9 and 5 million respectively.
   
      Similar to the previous case, fifteen snapshots of the agent configurations have been shown in Fig. 7 for $L = 256$. 
   These snapshots are taken
   at long times after the jumps to the ordered states have taken place. Here a number of bands of different orientations and 
   wrappings have been observed. Compared to $L = 128$ we find here a new type of stationary state where two different sets of bands
   cross one another. For example, in case of $\eta = 1.800$, we have exhibited a snapshot where a single horizontal band $W(0,1)$
   with $\phi = \pi/2$ crosses a single diagonal band $W(1,1)$ with $\phi = \pi/4$. Both bands are stable and move along two different
   directions, separated by $\pi/4$. In general, for such crossed bands, the order parameter takes slightly smaller values.

      In Fig. 8 the time averaged value of the order parameter $\Omega(\eta,L)$ has been plotted against
   $\eta$. The sharp fall in the order parameter takes place at $\eta_c(L) \approx 2.350$. Beyond this
   value of $\eta > \eta_c(L)$, order parameter is nearly zero. For $\eta < \eta_c(L)$ four sets of data
   points are plotted which correspond to four different band structures as explained in the figure caption.

\subsection {System size L = 512}

      The bands become most clearly visible for the system size $L = 512$. Because of the choice of random orientation 
   angles of the velocity vectors, the initial state is disordered with vanishingly small value of the order parameter. 
   Beyond the critical noise value $\eta_c = 2.368$ the order parameter in the stationary state assumes a very small 
   value. However, when the noise is reduced, $\Omega(\eta_c,L)$ jumps discontinuously at $\eta_c$ to a finite value. 
   In Fig. 9 the instantaneous value of the order parameter $\Omega(t,\eta,L)$ has been plotted against time for long 
   durations. For $\eta$ = 2.350 and 2.360 discontinuous jumps to the ordered state are observed after approximately 9 
   and 21 million time steps. In the bottom curve, $\eta = 2.370$ has been used and no such jump has been observed within 
   the entire duration of observation of 35 million time steps. 

      Therefore in general, the long time stationary states of all states for $\eta < \eta_c(L) = 2.368$ have been 
   observed to be characterized by the presence of bands of high density directed agents. As the value of the noise 
   strength $\eta$ is systematically decreased different types of bands appear in the stationary states. In Fig. 10 we 
   have exhibited twenty snapshots of agent locations in these stationary states of $L$ = 512. We have observed single and 
   multiple bands, diagonal and non-diagonally oriented bands, crossed bands meeting perpendicular to one another or 
   at an angle, and also bands which wrap the system multiple times. For $0 < \eta < 1$ the clearly visible band 
   structure has been missing till our time of observation of 50 million time steps. In Table III we describe the band 
   structures of these twenty stationary states. The value of $\eta$ has been mentioned below 
   every plot.

      In Fig. 11 we have plotted $\Omega(\eta,L)$ against $\eta$. Here also, in the ordered state the points for 
   different types of bands form different groups. Within one group the band pattern is the same for all values 
   of $\eta$ but the value of $\Omega(\eta,L)$ increases linearly on decreasing $\eta$. Six such group of points 
   have been shown in Fig. 11 and the corresponding wrapping numbers have been mentioned in the caption. 
   
      Before we summarize we like to emphasize two points. First, we try to visualize the motion of crossed bands
   where the direction of motion of each band is perpendicular to its sharp edge. As the system evolves the bands 
   move through each other, but because of the periodic boundary condition, they always remain entangled and cannot 
   be detached from each other. In principle, the simultaneous appearance of three or more different sets of bands 
   crossed among themselves may be quite possible, may be for larger system sizes, but we have not observed them. 
   In Fig. 12 we present four snapshots of the crossed bands at a small interval of 100 time units to exhibit their 
   motion. Two parallel $W(1,0)$ bands are crossed by one $W(1,1)$ band and all three bands move perpendicular to their
   front edges.
      
\begin{figure*}[t]
\begin {tabular}{cccc}
\includegraphics[width=4.0cm]{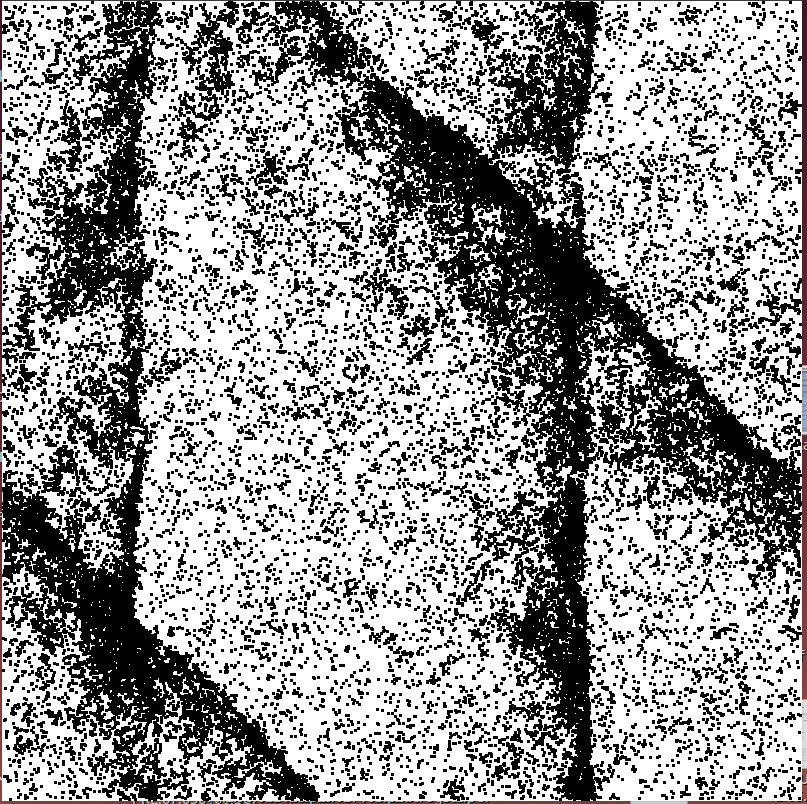} & \includegraphics[width=4.0cm]{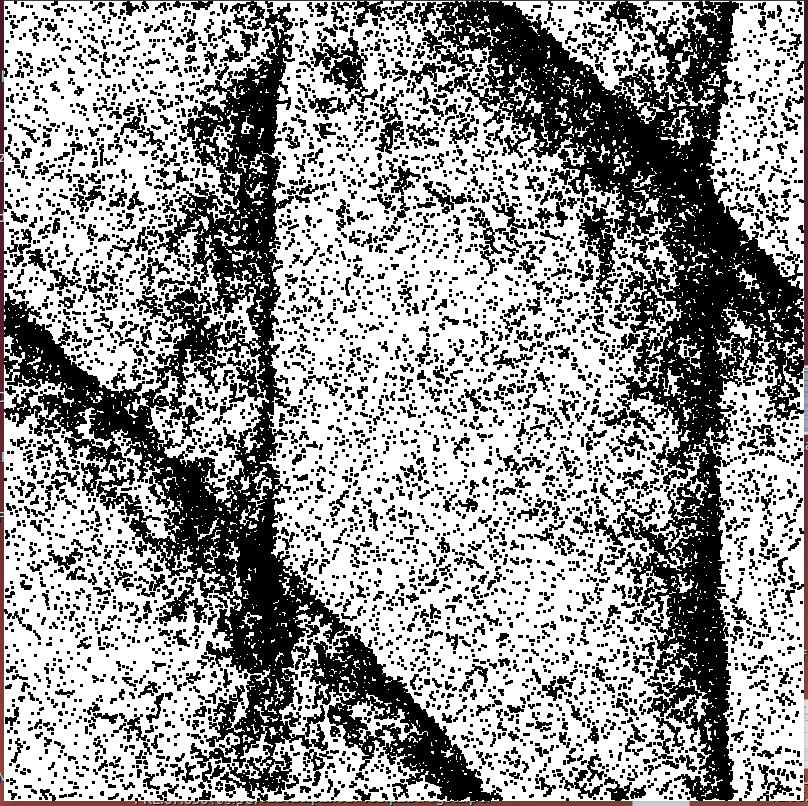} &
\includegraphics[width=4.0cm]{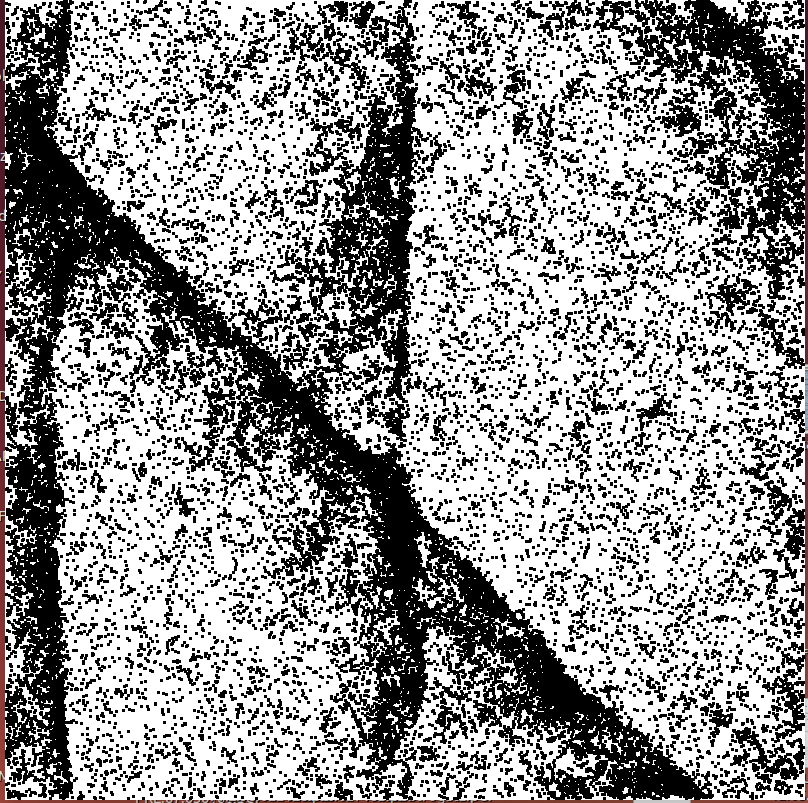} & \includegraphics[width=4.0cm]{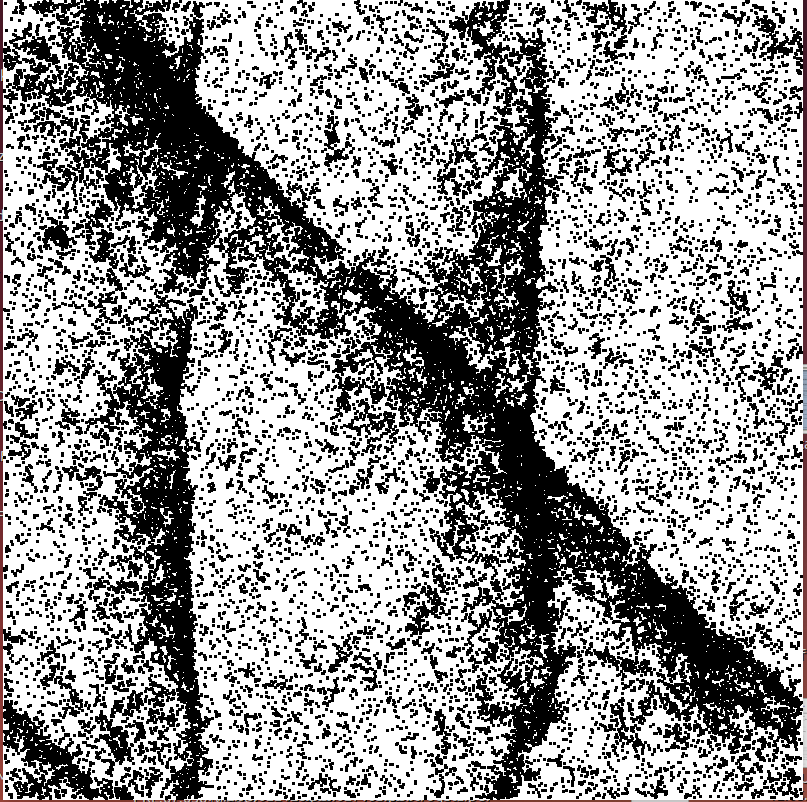} \\
\end {tabular}
\caption{
Crossed band configurations at the stationary state for a system size $L$ = 256,
$\eta$ = 1.8, and with density $\rho = 1$ of the agents. Starting from an arbitrary
initial configuration, the system arrives at the stationary having two W(1,0) bands 
and one W(1,1) band. Four snapshots have been shown at the interval of 100 time steps
from left to right. These figures exhibits that W(1,0) bands are moving to the right
and the W(1,1) bond is moving towards the top-right direction.
}
\label {fig12}
\end{figure*}
      
      Secondly, in Fig. 13 we present four snapshots of the stationary state bands starting from four different initial 
   configurations of positions and velocities of the agents, the strength $\eta$ of the noise being maintained 
   the same. We have obtained different wrapping patterns in each case. It may be quite possible that there are 
   more patterns, and starting from a random initial configuration the system may evolve to any one of the band
   patterns in the stationary state. We also believe that the probabilities of occurrence of these steady states 
   are likely to be distinct and possibly function of $\eta$. However, to estimate this probability distribution 
   numerically needs high value of computational resources and unfortunately we could not afford that. Figs. 4, 7 and 
   10 have different system sizes, namely $L$ = 128, 256, and 512. We observed that larger the system size, more 
   and more complicated bands with larger values of $m$ and $n$ appear in the stationary states.
   
\begin{figure*}[t]
\begin {tabular}{cccc}
\includegraphics[width=4.0cm]{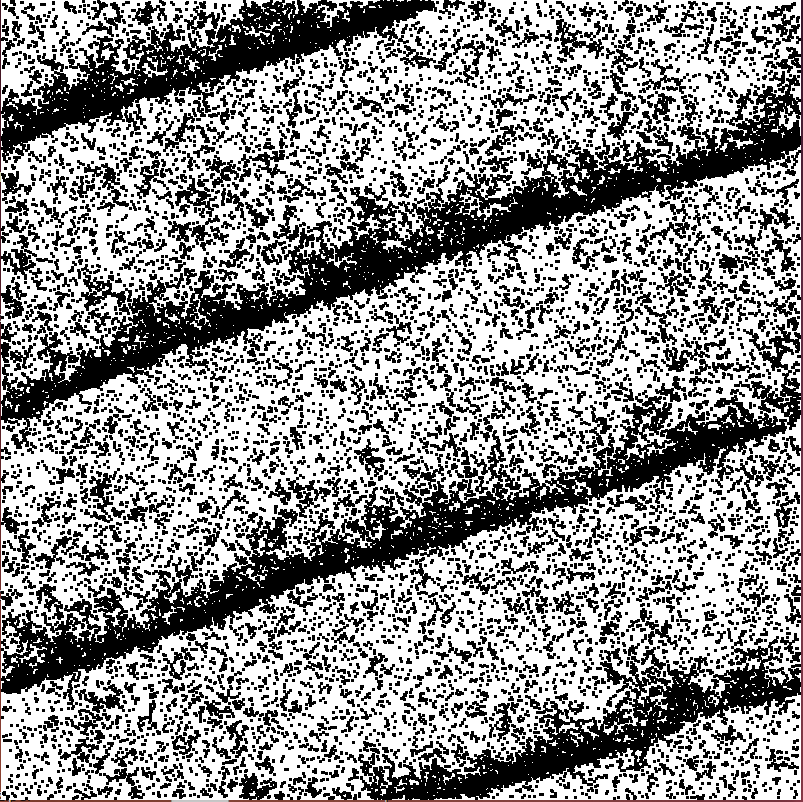} & \includegraphics[width=4.0cm]{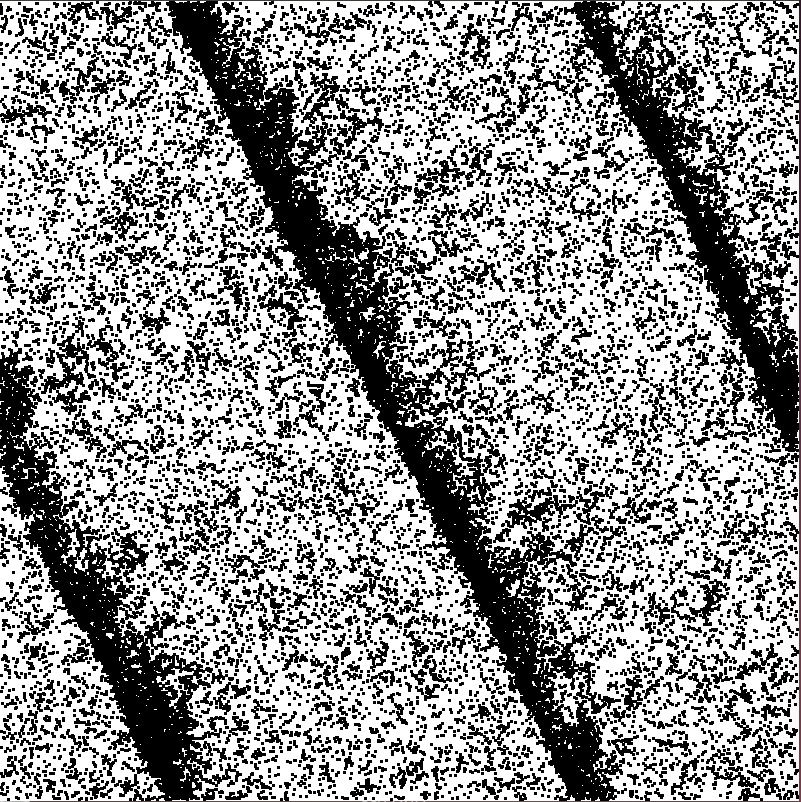} &
\includegraphics[width=4.0cm]{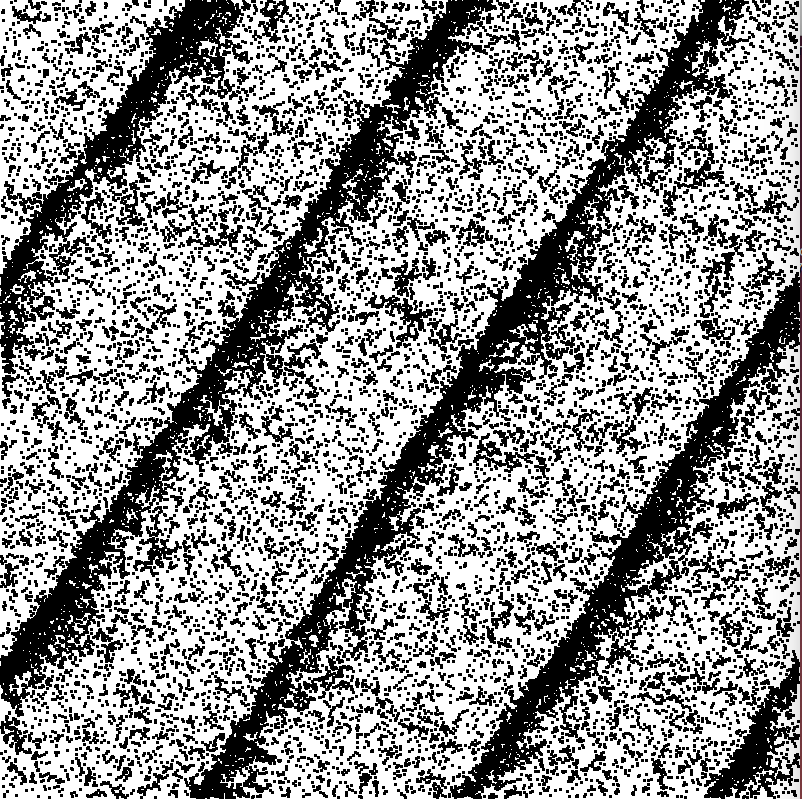} & \includegraphics[width=4.0cm]{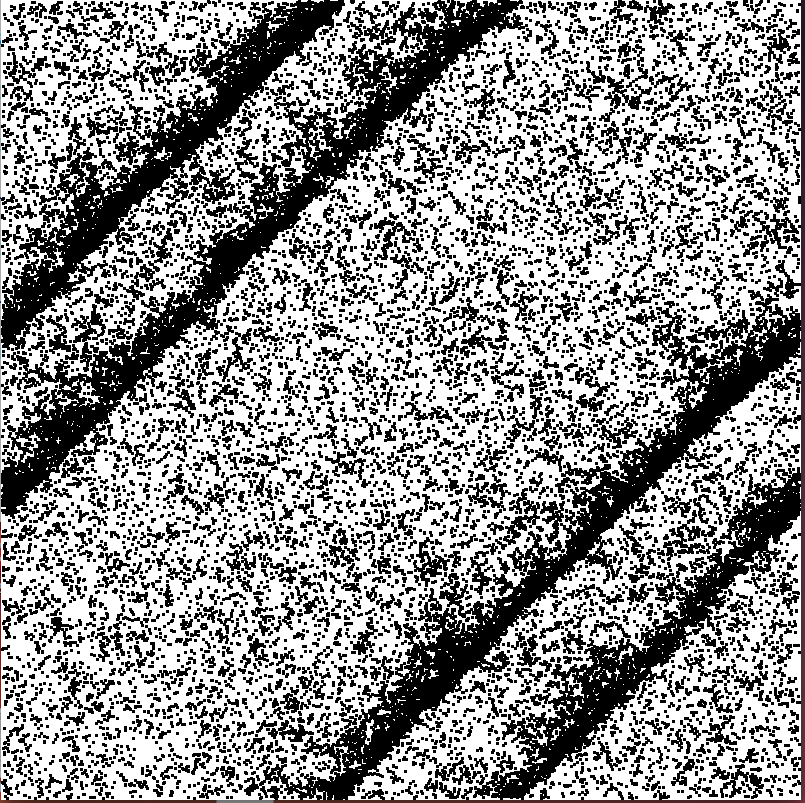} \\
\end {tabular}
\caption{
Different band configurations at the stationary state for a system size $L$ = 256,
$\eta$ = 2.2, and with density $\rho = 1$ of the agents. Different sets of 
initial positions of the agents and different directions for their initial velocities
have been used to simulate these systems. The stationary state band configurations are 
quite different e.g., starting from the left we get $W(1,3)$, $W(2,1)$, $W(3,2)$, and $2W(1,1)$
bands.
}
\label {fig13}
\end{figure*}

\section {4. Summary}

      To summarize, we have studied the Vicsek model of collective motion with quenched range of interaction. For studying
   the two dimensional version of the model with scalar noise, the underlying plane has been divided into non-overlapping 
   square shaped neighborhoods. All agents residing within a certain square cell at a certain time are mutual neighbors of 
   one another. Direction of the resultant of their velocity vectors is assigned to all agents which are then topped up by
   applying random noise. In a microscopic description of the model it has been argued that the agents in two adjacent cells 
   having similar velocity directions feel certain cohesiveness and therefore continue their motion in the adjacent cells.
   This cohesiveness property of moving together along the same direction is the original cause for the formation of band
   structures in the models of collective motion in the framework of the Vicsek model. Within a band, the agents are 
   correlated. As the noise decreases the sub-critical regime, the correlation becomes stronger. Consequently the system 
   becomes more ordered which is reflected in the non-trivial shaped bands of longer lengths, larger widths, different 
   orientations and wrapping numbers. We have formulated a detailed prescription for the characterization of such bands. 
   Starting from the completely disordered regime, as the strength of the noise is tuned down systematically, the most simple
   band parallel to the edges pops up abruptly, indicating a discontinuous transition similar to the Vicsek model. 

      By introducing the quenched range of interaction we have reduced, the individual freedom of the agents. In a way,
   this can be looked upon as feeding more correlation among the agents. It seems likely, that because of this extra correlation
   many different shaped bands show up in our model, even for the system size $L = 512$. Appearance of similar bands may be 
   plausible even in the original Vicsek model for the larger system sizes and with smaller noise.
   
      We thank the referee for pointing out that wrapping patterns of occupied sites are observed in the model of percolation
   as well \cite {Pinson}. In a similar way a pair of numbers $(m,n)$ have been used to characterize the patterns there as well.

\end {document}